\pdfoutput=1    
\documentclass[format=acmsmall, authorversion=true, review=false, timestamp=true, screen=true, nonacm=true]{acmart}

\renewcommand\footnotetextcopyrightpermission[1]{} 
\usepackage{xspace}
\usepackage{booktabs} 
\usepackage[ruled]{algorithm2e} 
\usepackage[caption=false]{subfig}

\usepackage{subfiles}
\usepackage{rotating}
\usepackage{xcolor,colortbl}

\definecolor{shadeOne}{rgb}{0.9,0.95,0.95}
\definecolor{shadeTwo}{rgb}{0.99,0.99,0.99}

\newcommand\etalii				{\xspace\textit{et al.}\xspace}
\newcommand{\commentt}[1]{}

\SetAlFnt{\small}
\SetAlCapFnt{\small}
\SetAlCapNameFnt{\small}
\SetAlCapHSkip{0pt}
\IncMargin{-\parindent}

\setcopyright{rightsretained}

\acmDOI{}

\begin{document}
\title{Exploiting Errors for Efficiency: A Survey from Circuits to Algorithms%
}

\author{Phillip Stanley-Marbell}
\orcid{0000-0001-7752-2083}
\affiliation{%
  \institution{University of Cambridge}
}
 \email{(phillip.stanley-marbell@eng.cam.ac.uk)}

\author{Armin Alaghi}
\affiliation{%
 \institution{University of Washington}
}

\author{Michael Carbin}
\affiliation{%
 \institution{Massachusetts Institute of Technology}
}

\author{Eva Darulova}
\affiliation{%
  \institution{Max Planck Institute for Software Systems}
}

\author{Lara Dolecek}
\affiliation{%
 \institution{University of California at Los Angeles}
}

\author{Andreas Gerstlauer}
\affiliation{%
 \institution{The University of Texas at Austin}
}

\author{Ghayoor Gillani}
\affiliation{%
 \institution{University of Twente}
}

\author{Djordje Jevdjic}
\affiliation{%
 \institution{National University of Singapore} 
}

\author{Thierry Moreau}
\affiliation{%
 \institution{University of Washington}
}

\author{Mattia Cacciotti}
\affiliation{%
 \institution{\'Ecole Polytechnique F\'ed\'erale de Lausanne}
}

\author{Alexandros Daglis}
\affiliation{%
 \institution{\'Ecole Polytechnique F\'ed\'erale de Lausanne}
}

\author{Natalie Enright Jerger}
\affiliation{%
 \institution{University of Toronto}
}

\author{Babak Falsafi}
\affiliation{%
 \institution{\'Ecole Polytechnique F\'ed\'erale de Lausanne}
}

\author{Sasa Misailovic}
\affiliation{%
 \institution{University of Illinois at Urbana-Champaign}
}

\author{Adrian Sampson}
\affiliation{%
 \institution{Cornell University}
}

\author{Damien Zufferey}
\affiliation{%
 \institution{Max Planck Institute for Software Systems}
}

%
%
%
\begin{abstract}
When a computational task tolerates a relaxation of its specification
or when an algorithm tolerates the effects of noise in its execution,
hardware, programming languages, and system software can trade
deviations from correct behavior for lower resource usage.  We
present, for the first time, a synthesis of research results on
computing systems that only make as many errors as their users can
tolerate, from across the disciplines of computer aided design of
circuits, digital system design, computer architecture, programming
languages, operating systems, and information theory.

Rather than over-provisioning resources at each layer to avoid
errors, it can be more efficient to exploit the masking of errors
occurring at one layer which can prevent them from propagating to
a higher layer. We survey tradeoffs for individual layers of computing
systems from the circuit level to the operating system level and
illustrate the potential benefits of end-to-end approaches using
two illustrative examples. To tie together the survey, we present
a consistent formalization of terminology, across the layers, which
does not significantly deviate from the terminology traditionally
used by research communities in their layer of focus.
\end{abstract}

%
%

%
%

\keywords{Approximate computing, error efficiency, cross-layer, end-to-end.}


\maketitle

\renewcommand{\shortauthors}{Stanley-Marbell\etalii}
\newcounter{TodoCounter} 


\vspace{-0.05in}
\section{Introduction}
\vspace{-0.05in}

Computing systems solve specific computational problems by transforming
an algorithm's inputs to its outputs. This, as well as counteracting
the effects of noise in the underlying hardware
substrate~\cite{bennett1985fundamental, keyes1985makes, Shannon:59},
requires resources such as time, energy, or hardware real-estate.  Resource
efficiency is becoming an increasingly important challenge, especially
due to the pervasiveness of computing systems and the
diminishing returns from performance improvements of process
technology scaling~\cite{amirtharajah2004micropower,
Palem:2005:EAC:1079838.1080003, Breuer:2005:MAI:1099541.1100164}.
Computing systems are reaching
the fundamental limits of the energy required for fully reliable
computation~\cite{bennett1985fundamental, Markov14}.

At the same time, many important applications have nondeterministic
specifications or are robust to noise in their execution. They thus
do not necessarily require fully reliable computing systems and
their resource consumption can be reduced.  For instance, many
applications processing physical-world signals often have multiple
acceptable outputs for a large part of their input domain.  Furthermore,
all measurements of analog signals have some amount of measurement
uncertainty or noise, and digital signal representations necessarily
introduce quantization noise. It is therefore impossible to perform exact
computation on data resulting from real-world, physical signals.

These observations about the fundamental limits of computation and
the possibility of trading correctness for resource usage 
have always been implicit in computing systems design dating back
to the ENIAC~\cite{VonNeumann:problogic}, but have seen
renewed interest in the last decade. This interest has focused
on techniques to trade precision, accuracy, and reliability for
reduced resource usage in hardware. These recent efforts
harness nondeterminism and take advantage of application
tolerance to coarser discretization in time or value (i.e., precision
or sampling rate), to obtain significant resource savings for 
an acceptable reduction in accuracy and reliability.
These techniques have been referred to in the research literature as
\textit{approximate computing} and include:
\begin{itemize}
\item   Programming languages to specify computational problem and algorithm
	nondeterminism.

\item   Compilation techniques to transform specifications which
	expose nondeterminism or flexibility, into concrete
	deterministic implementations.

\item   Hardware architectures that can exploit nondeterminism
	exposed at the software layer, or which expose hardware
	correctness versus resource usage tradeoffs to the
	layers above.

\item   New devices and circuits to implement architectures that
	exploit or expose nondeterminism and correctness
	versus resource usage tradeoffs.
\end{itemize}
In the same way that computing systems that only use as much energy
as is necessary are referred to as being \emph{energy-efficient},
we can refer to the computing systems investigated in this survey
as being \emph{error-efficient}: they only make as many errors as
their users can tolerate~\cite{EDA-049}.

\vspace{-0.05in}
\subsection{Related surveys}
\vspace{-0.05in}
%
%
%
This survey explores techniques for hardware and software systems
in which the system's designers or its users are willing to trade
lower resource usage for increased occurrence of deviations from
correctness. These deviations from correctness may occur within an
individual layer of the system stack, or they may occur in the
context of an end-to-end computing system.  Correspondingly,
techniques have been developed for all of the systems layers: for
the transistor-, gate-, circuit- or microarchitecture-, architecture-,
language-/runtime-, and system-/software-level.  Multiple surveys
of approximate computing (and related techniques) exist in the
literature~\cite{cct:survey, Xu2016, Mittal2016, MoreauTaxonomy2018,
Aponte2018, shafique2016invited}. This survey provides the first holistic
overview of fundamental limits
of computation in the presence of noise, probabilistic computing,
stochastic computing, and voltage overscaling \emph{across the
computing system stack}. 
Such a holistic consideration is important in order to make these 
techniques useful for real systems and to enable increased resource 
savings.  At the same time, it requires collaborations between 
different areas and communities with often differing terminology.

\vspace{-0.05in}
\subsection{Contributions and outline}
\vspace{-0.05in}
This survey presents:
\begin{itemize}
\item   \textbf{A cross-disciplinary overview} of research on
	correctness versus resource usage tradeoffs spanning the
	hardware abstractions and disciplines of: transistors,
	circuits, microarchitecture and architecture, programming languages, and operating
	systems.

\item   \textbf{An overview of existing uses of quality versus
	resource usage tradeoffs} across application domains and
	\textbf{examples of two end-to-end applications}
    (Section~\ref{section:applications}
	and Section~\ref{section:examples}).

\item   \textbf{Terminology} for describing
	resource usage versus correctness tradeoffs of computing
	systems that interact with the physical world
	that is consistent with
	existing widely-used terminology but which at the same time
	provides a coherent way to discuss these tradeoffs across
	domains of expertise (Section~\ref{section:terminology}).

\item   \textbf{Detailed discussions of the state of the art}
	across the layers of system implementation stack from
	circuits, to microarchitecture and architecture, to the
	programming language and operating system layers of abstraction
	(Section~\ref{section:transistorGateAndCircuitLevel}\,--\,Section~\ref{section:OsLevel}).

\item   \textbf{A taxonomy} tying
	together the ideas introduced in the survey (Section~\ref{section:taxonomy}).

\item   \textbf{A discussion of limits of computation
	in the presence of noise}.
%
%
%
	(Section~\ref{section:fundamentalLimitsOfNoisyComputation}).

\item   \textbf{A set of open challenges}
	across the layers of
	abstraction (Section~\ref{section:challenges}).
\end{itemize}

%
%
%

%
%
%

%
%
%

\vspace{-0.05in}
\section{Existing Quality versus Resource Usage Tradeoffs}
\label{section:applications}
\vspace{-0.05in}
The idea of trading quality for resources and efficiency is inherent
to all computing domains.  Several research communities have developed
techniques to exploit tolerance of applications to noise, errors,
and approximations to improve the reliability or efficiency of
software and hardware systems.  In the same way that there have
always been attempts to make hardware and software more tolerant
to faults independent of specific research on fault-tolerant
computing, there has also always been a pervasive use of techniques
for approximation (e.g., Taylor series expansions) independent of
recent interest in approximate computing. The following highlights
some of these efforts across application domains.

\vspace{-0.05in}
\subsection{Scientific computing}
\vspace{-0.05in}
Scientific computing can be defined as ``the collection of tools, techniques, and
theories required to solve on a computer mathematical models of
problems in science and engineering''~\cite{Golub2014}. Most of
these models are real-valued, and exact analytical solutions rarely
exist or are costly to compute~\cite{Meerschaert2013,Constanda2017}.
As a result, numerical approximations and their associated
quality-efficiency tradeoffs have always been important in scientific
computing~\cite{Einarsson2005}.

These numerical approximations are introduced at different levels
of abstraction. Because the real-world is too complex to be represented
exactly, practical considerations require resorting to models,
incurring modeling errors~\cite{Meerschaert2013}. Even with a model
in hand, analytical solutions may not exist and numerical solutions
are needed to approximate the exact
answers~\cite{Dahlquist2008,Burden2015}, introducing further
deviations from the expected result.  And finally, most models are
real-valued and thus have to be approximated by finite-precision
arithmetic, adding roundoff errors~\cite{Higham2002}.

Roundoff errors can be bounded to some extent automatically using
techniques such as interval arithmetic~\cite{Rump2010}.  Dealing
with most of the errors introduced by modeling, numerical approximation,
and finite-precision arithmetic, is rarely automated by software
tools. The state of the art in dealing with modeling and numerical
errors often requires manual intervention of the programmer or
domain expert and is typically on a per-application basis. Because
of the resulting complexity of the error analysis, the resulting
error bounds are often only asymptotic.
%
%

\vspace{-0.05in}
\subsection{Embedded, digital signal processing, and multimedia systems}
\label{section:02:embedded-and-multimedia}
\vspace{-0.05in}
Many computing systems that interact with the physical world or
which process data gathered from it, have high computational demands
under tightly-constrained resources. These systems, which include
many embedded and multi-media systems, must often process noisy
inputs and must trade fidelity of their outputs for lower resource
usage. Because they are designed to process data from noisy inputs,
such as from sensors that convert from an analog signal into a
digital representation, these applications are often designed to
be resilient to errors or noise in their inputs~\cite{189934}.

Several pioneering research efforts investigated trading precision
and accuracy for signal processing
performance~\cite{amirtharajah2004micropower} and exploiting the
tolerance of signal processing algorithms to noise~\cite{Shanbhag:2002,
HShanbhag:ANT}. When the outputs of such systems are destined for
human consumption (e.g., audio and video), common use cases can
often tolerate some amount of noise in their I/O
interfaces~\cite{Stanley-Marbell:2016:CSP:2901318.2901347, 1804.02317,
Stanley-Marbell:2016:RSI:2897937.2898079, stanley2015efficiency,
hotchips16encoder}.

\vspace{-0.05in}
\subsection{Computer vision, augmented reality, and virtual reality}
\vspace{-0.05in}
Many applications in computer vision, augmented reality, and virtual
reality are compute-intensive. As a result, many of their algorithms
(e.g., stereo matching algorithms) have always been implemented
with quality versus efficiency tradeoffs in
mind~\cite{scharstein2002taxonomy, tombari2008classification,
gong2007performance}.  The implementations of these algorithms have
used techniques including fixed-point implementations of expensive
floating-point numerics~\cite{menant2014optimized} and algorithmic
approximations such as removing time-consuming backtracking
steps~\cite{brombergerexploiting} when implementing these algorithms
on FPGA accelerators.



\vspace{-0.05in}
\subsection{Communications and storage systems}
\vspace{-0.05in}
The techniques we survey often involve computation on noisy inputs
or data processing in the presence of noise in much the same way
research in communication systems and information theory considers
communication over a noisy channel.  As one recent example of work
that could be viewed as either traditional information theory and
communication systems research or approximate computing,
Huang\etalii~\cite{huang2015acoco} present a simple yet effective
coding scheme that uses a combination of a lossy source/channel 
coding to protect against hardware errors for iterative
statistical inference.

\vspace{-0.05in}
\subsection{Big data and database management systems}
\vspace{-0.05in}
Approximate query processing in the context of databases and big
data research leverages sampling-based techniques to trade correctness
of results for faster query processing.  Early work in this direction
investigated sampling from
databases~\cite{Olken:1990:RSD:1127110.1127118,olken1986simple}.
More recently, BlinkDB~\cite{agarwal2013blinkdb}, an approximate
query engine, allows users to trade accuracy for response time.
BlinkDB uses static optimizations to stratify data in a way that
permits dynamic sampling techniques at runtime to present results
annotated with meaningful error bars.  Other recent efforts include
Quickr~\cite{kandula2016quickr} and
ApproxHadoop~\cite{goiri2015approxhadoop}.

\vspace{-0.05in}
\subsection{Machine learning}
\label{section:02:machine-learning}
\vspace{-0.05in}
Machine learning techniques learn functions (or programs) from data
and this data is in practice either limited or noisy.  Larger
datasets typically lead to more accurate trained machine learning
models, but in practice training datasets must be limited due to
constraints on training time. As a result, many machine learning
methods must inherently grapple with the tradeoffs between efficiency
and correctness of the systems.

There are several techniques that allow machine learning systems
to trade accuracy for efficiency.  These techniques include
\textit{random dropout}~\cite{Srivastava2014}, which randomly removes
connections within a neural network to prevent overfitting during
training and to improve overall training accuracy.  Techniques such
as weight de-duplication and pruning~\cite{Han2015, Chen2015},
low-intensity convolution operators~\cite{Iandola2016, Howard2017},
network distillation~\cite{Romero2014}, and algorithmic approximations
based on matrix decomposition~\cite{Denton14,Kim2015} take advantage
of redundancy to minimize the parameter footprint of a given neural
network.  Weight quantization is yet another technique to reduce
computation and data movement costs in
hardware~\cite{Courbariaux2015,Jouppi2017}.


\vspace{-0.05in}
\section{Illustrative end-to-end examples}
\label{section:examples}
\vspace{-0.05in}
Many applications from the domains of signal processing and machine
learning have traditionally had to grapple with tradeoffs between
precision, accuracy, application output fidelity, performance, and
energy efficiency (see, e.g.,
Section~\ref{section:02:embedded-and-multimedia} and
Section~\ref{section:02:machine-learning}). Many of the techniques
applied in these domains have been reimagined in recent years, with
a greater willingness of system designers to explicitly trade reduced
quality for improved efficiency.

We discuss two applications from the signal processing and machine
learning domains: a pedometer and digit recognition. Using these
examples, we suggest ways in which resource usage versus correctness
tradeoffs can be applied across the layers of the hardware stack,
from sensors, over I/O, and to computation. We use these applications
to demonstrate how end-to-end resource usage could potentially be
improved even more when tradeoffs are exploited at more than one
layer of the system stack.


%
%
%
%

\begin{figure}
\centering
\includegraphics[trim=0cm 0cm 0cm 0cm, clip=true, angle=0, width=0.75\textwidth]{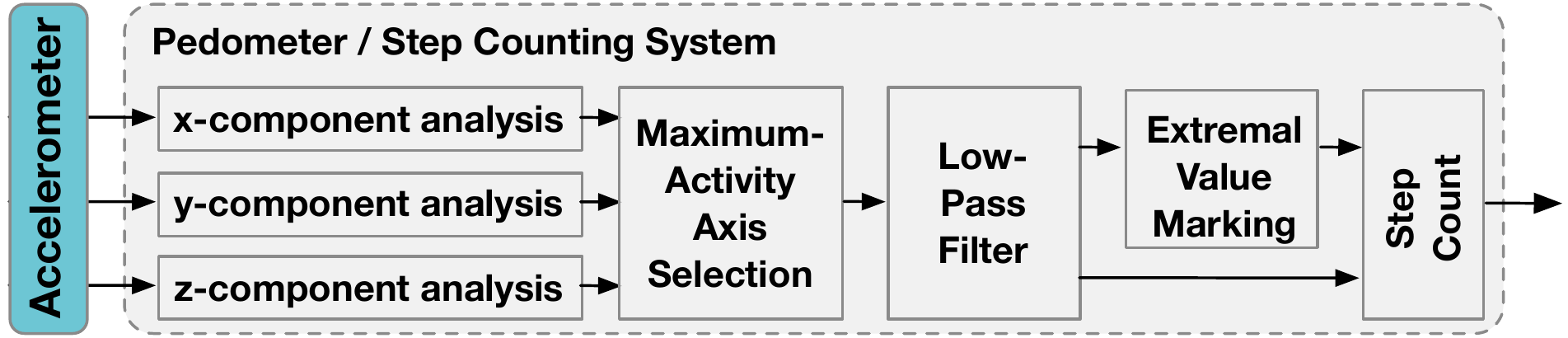}
\vspace{-0.1in}
\caption{The block diagram of one canonical pedometer application implementation.}
\label{fig:example:pedometer}
\vspace{-0.2in}
\end{figure}

\vspace{-0.05in}
\subsection{Example: a pedometer application}
\label{example:pedometer}
\vspace{-0.05in}
Applications which process data measured from the physical world
must often contend with noisy inputs. Signals such as temperature,
motion, etc., which are analyzed by such sensor-driven systems, are
usually the result of multiple interacting phenomena which measurement
equipment or sensors can rarely isolate. At the same time, the
results of these sensor signal processing applications may not have
a rigid reference for correctness.  This combination of input noise
and output flexibility leads to many sensor signal processing
applications having tradeoffs between correctness and resource
usage.

One concrete example of such an application is a pedometer (step
counter). Modern pedometers typically use data from 3-axis
accelerometers to determine the number of steps taken during a given
period of time.  Even when a pedometer's wearer is nominally
motionless, these accelerometers will detect some distribution of
(noisy) measured acceleration values.  At the same time, small
errors in the step count reported by a pedometer are often
inconsequential and therefore acceptable.


Figure~\ref{fig:example:pedometer} shows a block diagram for an
implementation of one popular
approach~\cite{AnalogDevices:pedometerAppnote}.  Our implementation
takes as input 3-axis accelerometer data and returns a step count
for time windows of 500\,ms.  The pedometer algorithm first selects
the accelerometer axis with the maximum peak-to-peak variation (the
\textit{maximum activity axis selection} block in
Figure~\ref{fig:example:pedometer}). The algorithm uses the selections
to create a new composite sequence of accelerometer samples.  Next,
the pedometer algorithm performs low-pass filtering, and then, for
each 500\,ms window, computes the maximum and minimum acceleration
values and the midpoint of this range (the \textit{extremal value
marking} block in Figure~\ref{fig:example:pedometer}).  Finally,
the algorithm counts how many times the low-pass filtered signal
crosses the per-window midpoints in one direction (e.g., from above
the midpoint to below it), and it reports this count as the number
of steps.


%
%
\begin{figure}[h]
\subfloat[All three axes of data (shown low-pass filtered).]{\includegraphics[trim=0.0cm 0.0cm 0.0cm 0.0cm, clip=true, angle=0, width=0.28\textwidth]{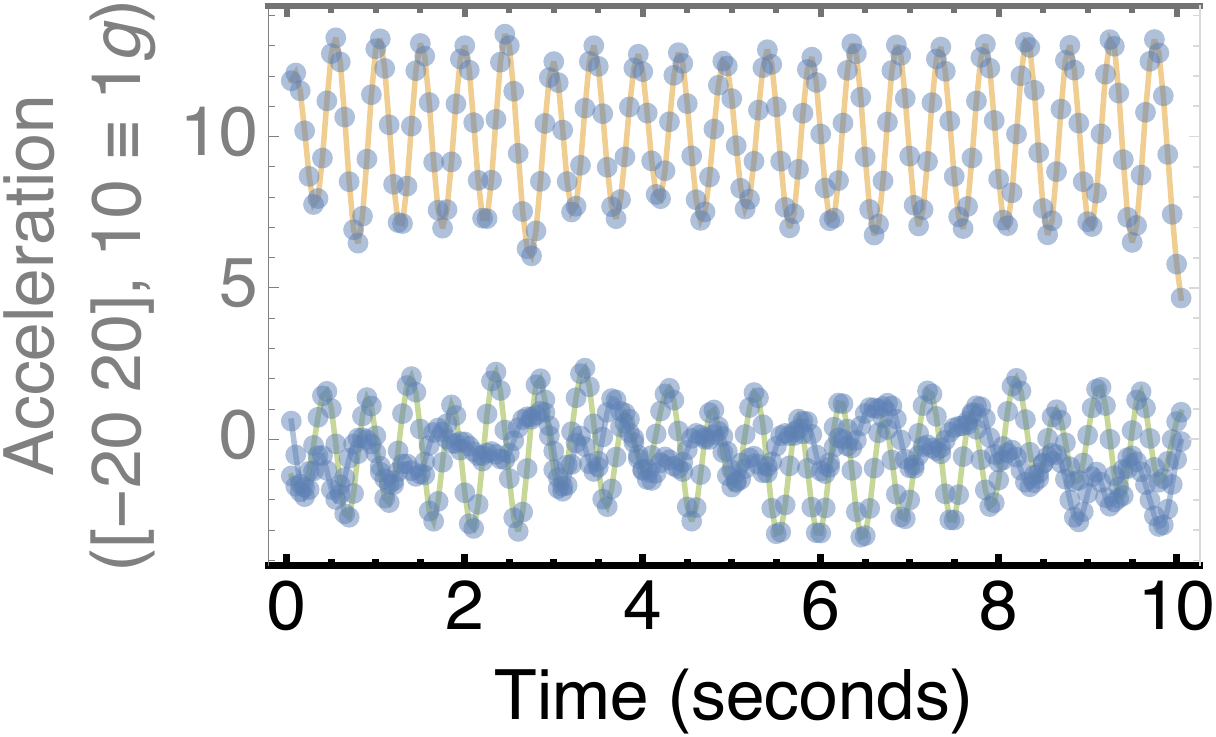}}\hspace{0.25in}
\subfloat[Maximum activity axes combined across the 500\,ms windows.]{\includegraphics[trim=0.0cm 0.0cm 0.0cm 0.0cm, clip=true, angle=0, width=0.28\textwidth]{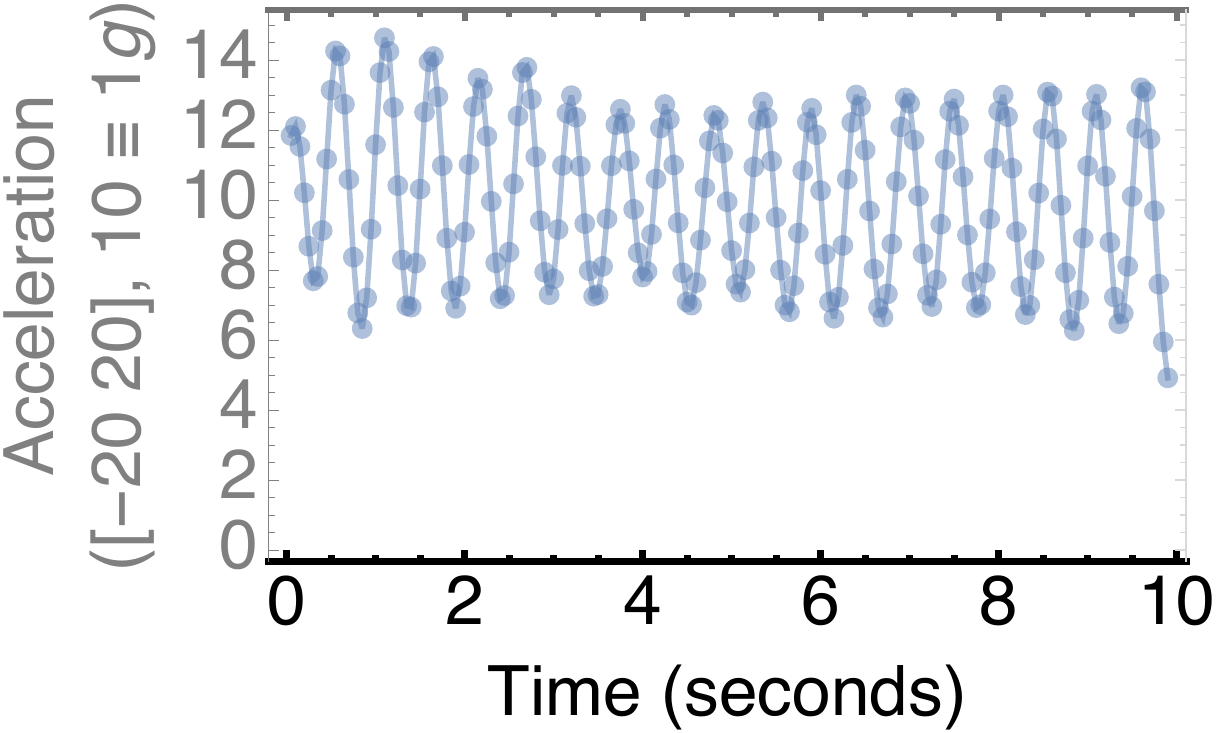}}\hspace{0.25in}
\subfloat[Extremal value marking of the maximum-activity axis data. Step count: 19.]{\includegraphics[trim=0.0cm 0.0cm 0.0cm 0.0cm, clip=true, angle=0, width=0.28\textwidth]{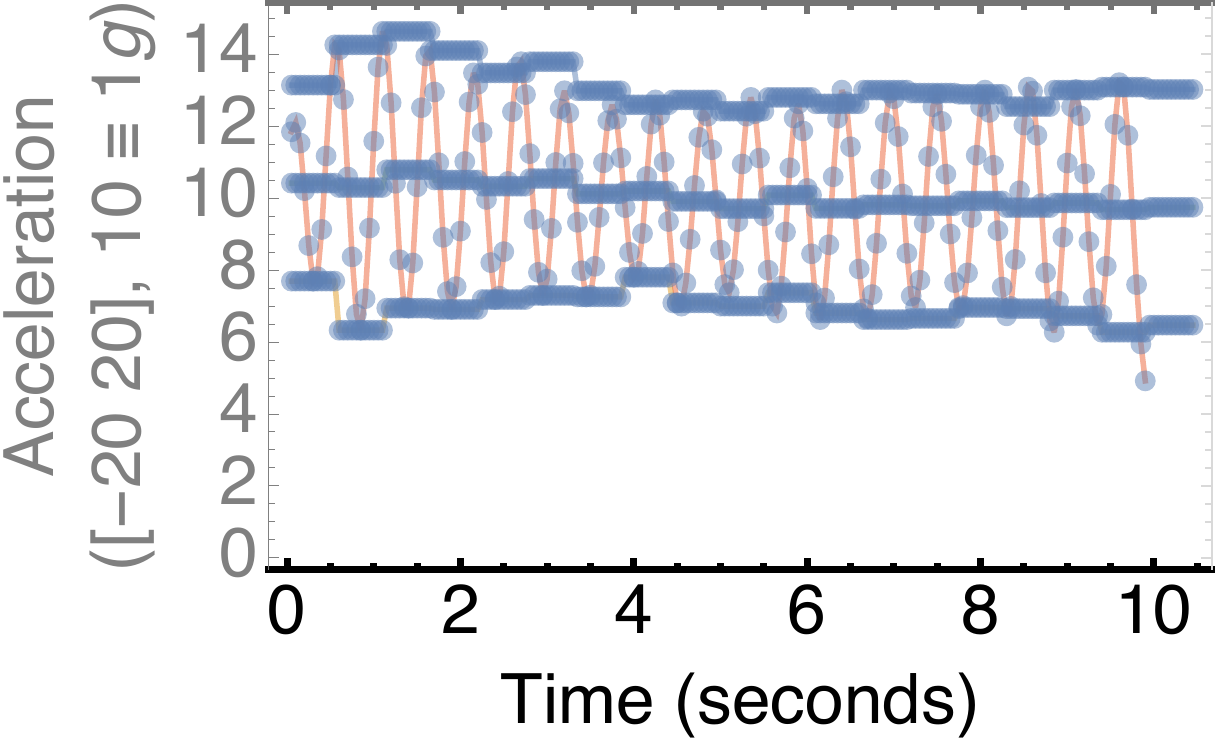}}\\
\subfloat[All three axes of data with added noise (shown low-pass filtered).]{\includegraphics[trim=0.0cm 0.0cm 0.0cm 0.0cm, clip=true, angle=0, width=0.28\textwidth]{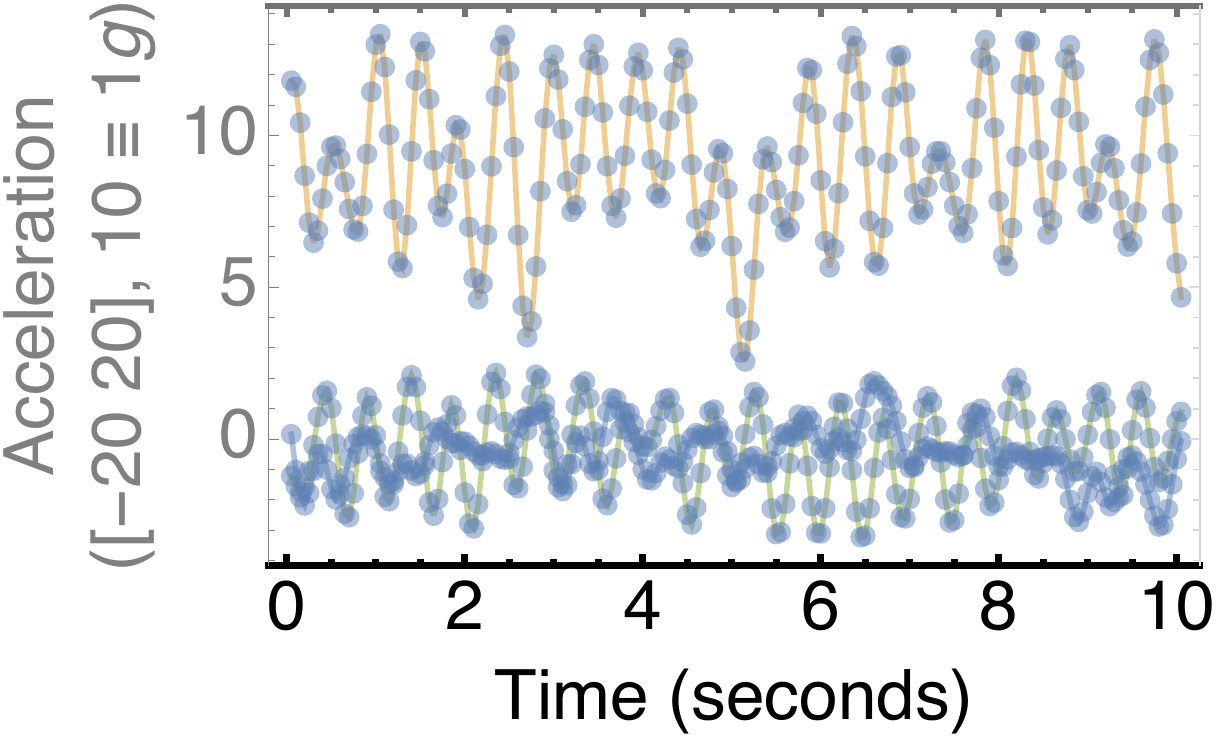}}\hspace{0.25in}
\subfloat[Maximum activity axes combined, for data with added noise.]{\includegraphics[trim=0.0cm 0.0cm 0.0cm 0.0cm, clip=true, angle=0, width=0.28\textwidth]{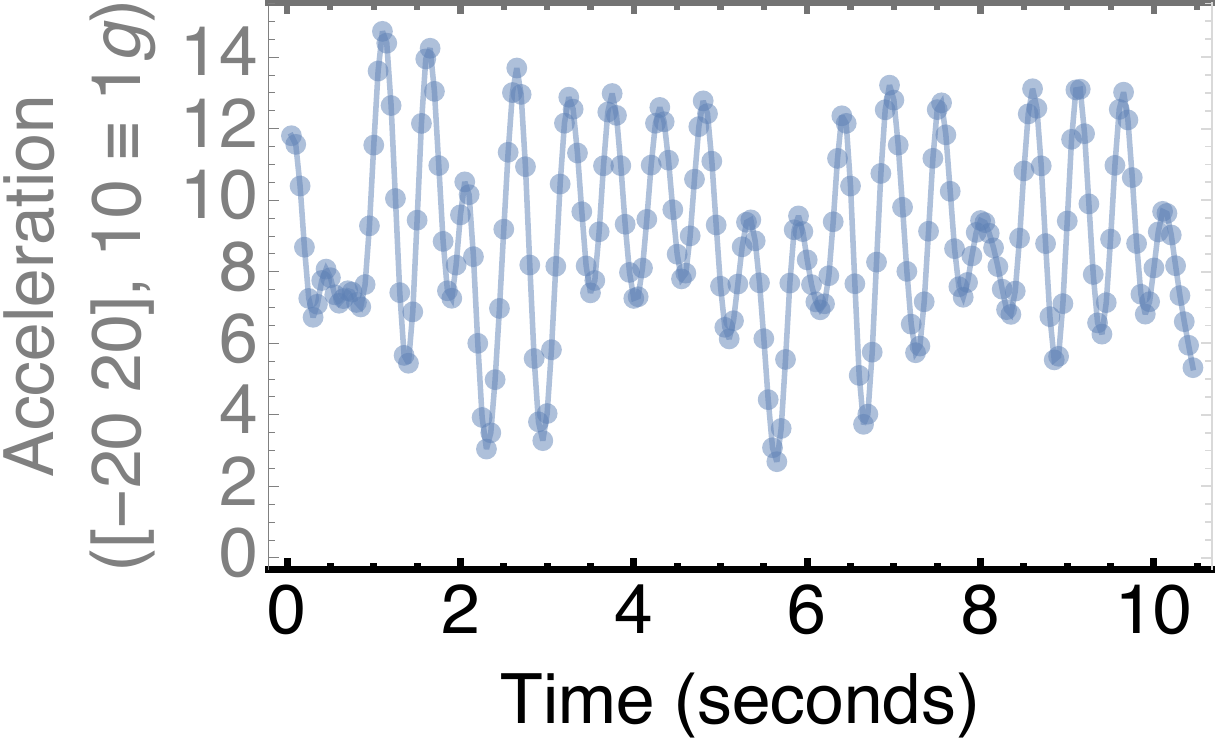}}\hspace{0.25in}
\subfloat[Extremal value marking for the data with added noise. Step count: 16.]{\includegraphics[trim=0.0cm 0.0cm 0.0cm 0.0cm, clip=true, angle=0, width=0.28\textwidth]{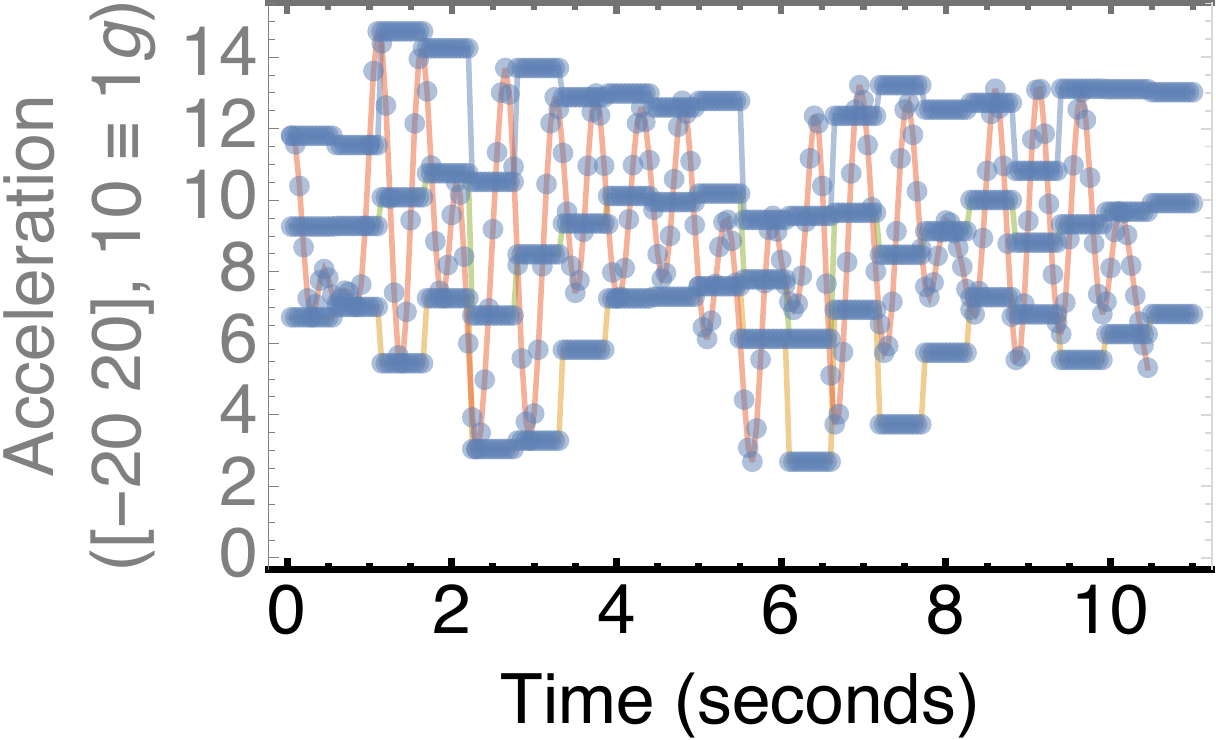}}
\vspace{-0.15in}
\caption{Intermediate stages of data from a pedometer application.}
\label{fig:example:pedometer-plots}
\vspace{-0.1in}
\end{figure}

Figure~\ref{fig:example:pedometer-plots}(a--c) show the progression
of a sequence of accelerometer samples through the stages of the
pedometer algorithm, which outputs a step count of 19 at the end.
Figure~\ref{fig:example:pedometer-plots}(d--f) show a modified
version of the data where we have replaced 5\% of the samples with
zeros to simulate intermittent failures at a sensor. Even though
the data in the final stage of the algorithm
(Figure~\ref{fig:example:pedometer-plots}(c) and
Figure~\ref{fig:example:pedometer-plots}(f)) looks qualitatively
different, the final output of the algorithm is relatively close
the noise-free output.
%
%
%

\textit{Applying individual tradeoffs.}
The hardware and system stack for a typical pedometer comprises
sensors (e.g., accelerometers), I/O links (e.g., SPI or I2C) between
those sensors and a processor, a runtime or embedded operating system,
the implementation of the pedometer algorithm, and a display.  A
system's designer may exploit the resource versus correctness tradeoffs
at each of these layers or components independently, using the
techniques surveyed in
Sections~\ref{section:transistorGateAndCircuitLevel}-\ref{section:OsLevel}
of this article. For example, a system designer could apply
Lax~\cite{189934} to sensors, VDBS encoding~\cite{1804.02317,
hotchips16encoder, Stanley-Marbell:2016:RSI:2897937.2898079} to the
I2C or SPI communication between sensors and a microcontroller, and
could ensure that the potentially inexact data does not affect the
overall safety of the application using EnerJ~\cite{Sampson2011}
or FlexJava~\cite{Park2015}.

\textit{Potential for end-to-end optimization.}
%
%
This survey argues for exploring the end-to-end combination of
techniques for trading correctness for efficiency, across the levels
of abstraction of computing systems.  Rather than treating each
layer of the hardware and system software stack as an independent
opportunity, this article argues that greater resource-correctness
tradeoffs are possible when the entire system stack is considered
end-to-end. For example, the insensitivity of the pedometer algorithm
to input noise highlighted in Figure~\ref{fig:example:pedometer-plots}
might be determined by program analyses. These analyses could in
turn be used to inform instruction selection for generated code as
well as determining sensor operating settings (e.g., sampling rate,
operating voltage, on-sensor averaging) and sensor I/O settings
(e.g., choices for the I/O encoding for the sensor samples as they
are transferred from a sensor).

%
%

\vspace{-0.05in}
\subsection{Example: digit recognition}
\label{section:MNIST}
\begin{figure}[t]
\centering
\subfloat[The MLP topology]{\includegraphics[trim=0.0cm 0.0cm 0.0cm 0.0cm, clip=true, angle=0, width=0.45\textwidth]{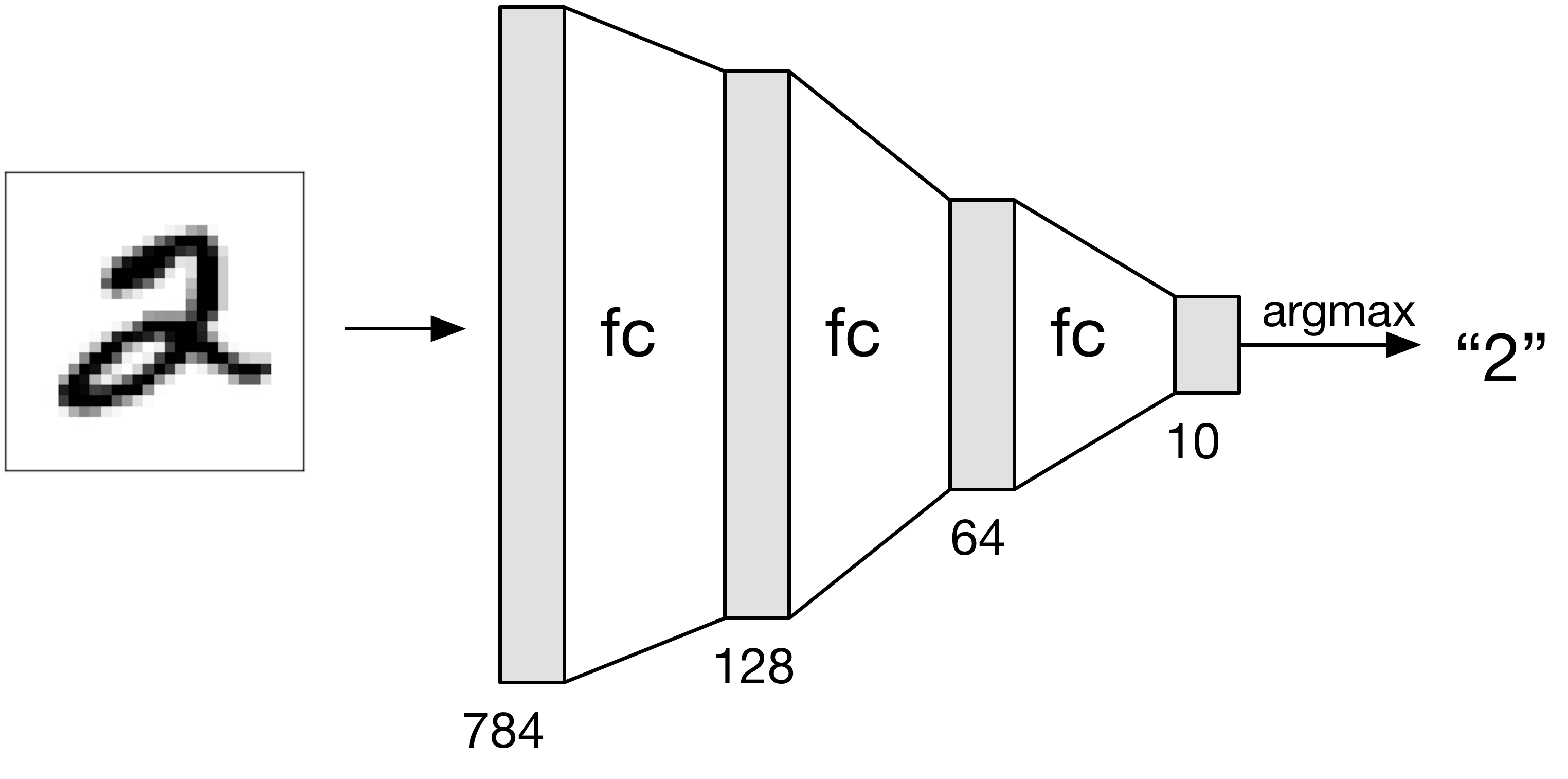}}
\subfloat[The effects of quantization]{\includegraphics[trim=0cm 0cm 0cm 0cm, clip=true, angle=0, width=0.45
\textwidth]{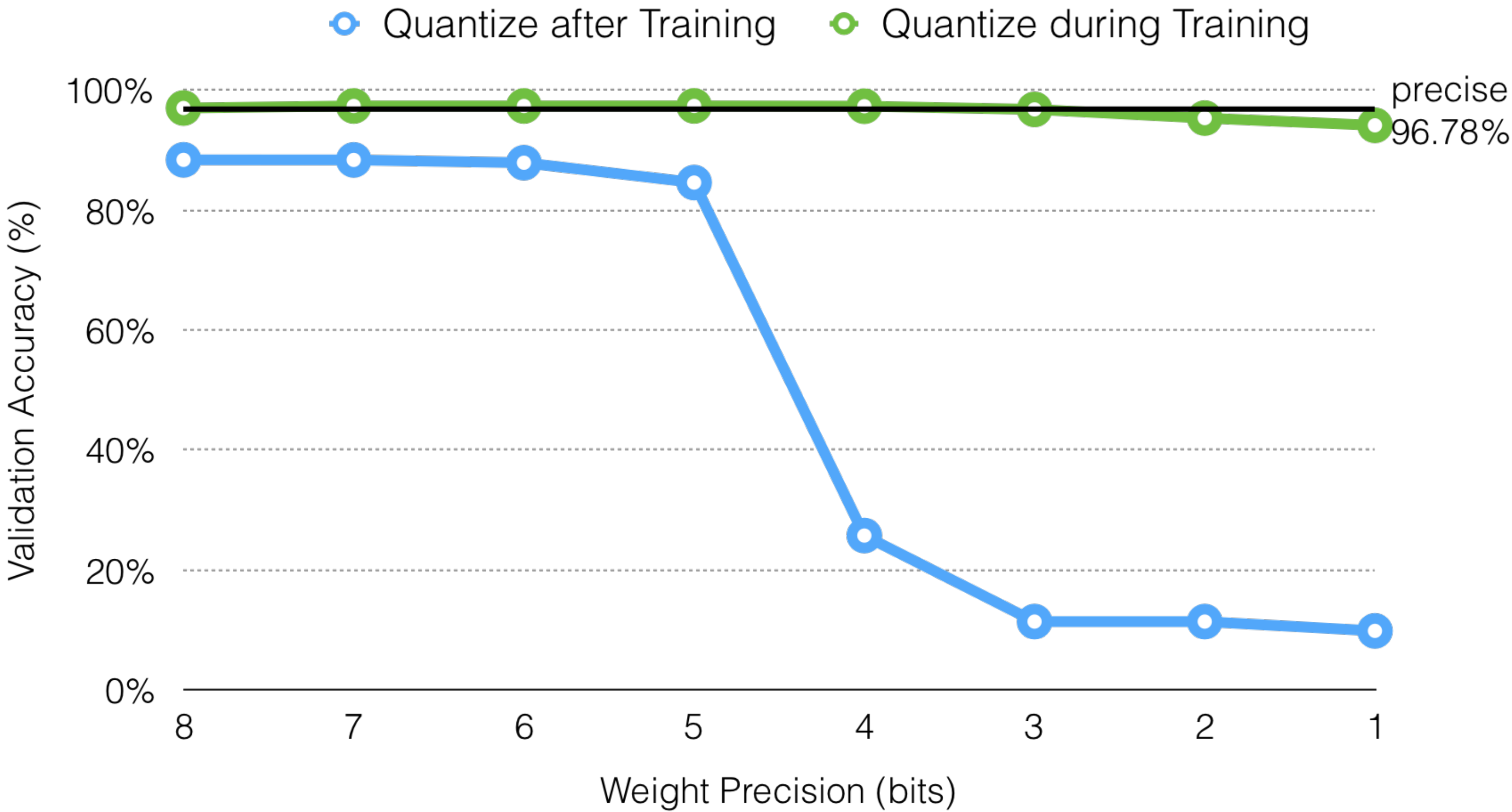}}
\vspace{-0.15in}
\caption{The multi-layer perceptron (MLP) trained on the MNIST dataset topology and quantization.}
\label{fig:intro:mnist_mlp}
\vspace{-0.15in}
\end{figure}

\vspace{-0.05in}
%
%
%
Digit recognition is the computational task of determining the
correct cardinal number corresponding to an image of a single
handwritten digit. One popular approach to implementing digit
recognition is using neural networks. In a neural network implementation
of digit recognition, pixel values from an input image of a standard
size (e.g., 28$\times$28 pixels) are fed into a neural network in
which the final layer encodes the digit value (a number between 0
and 9). Because of the compute-intensive nature of neural network
operators combined with the resilience to errors thanks to re-training
techniques~\cite{du2014leveraging}, neural networks are a compelling
target for resource versus correctness tradeoffs. Neural networks
for digit recognition are particularly interesting on devices where
energy efficiency is critical. Figure~\ref{fig:intro:mnist_mlp}(a)
shows a simple network architecture for performing handwritten digit
recognition. The network consists of three fully-connected layers
(labeled ``fc'' in the Figure). The input layer takes in a 28$\times$28
image with one node for each of the 784 pixels and the final output
layer has 10 nodes.

%
%


\textit{Applying individual tradeoffs.}
%
%
%
Figure~\ref{fig:intro:mnist_mlp} (b) shows the results for accuracy
of the neural network with quantization of weights starting with a
32-bit floating point baseline all the way down to a 1-bit weight.
The network is trained on the MNIST data set with quantization of
weights either during or after training.  The results show that as
long as re-training is applied, this neural network is extremely
tolerant even to aggressive quantization.


Quantization furthermore enables weight prunability and compressibility:
weights represented with fewer bits lead to fewer distinct values, and more
zero-valued weights.
This creates opportunities for sparse matrix compression~\cite{Han2016}, which can be directly implemented in hardware.

\begin{figure}[t]
\centering
\subfloat{{\includegraphics[trim=0cm 0cm 0cm 0cm, clip=true, angle=0, width=1\textwidth]{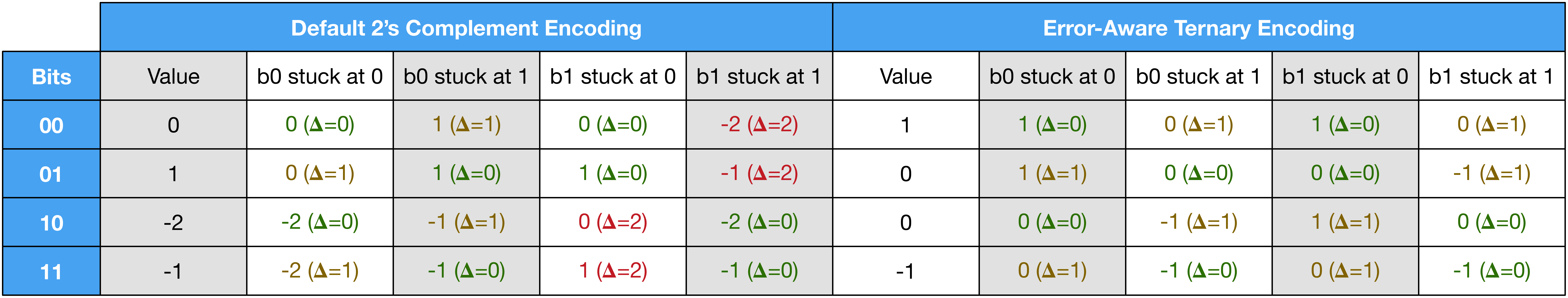}}}
\vspace{-0.15in}
\caption{Application aware codes can be critical in enabling robustness to errors in neural networks.
In the case of ternary neural networks, we can craft codes that guarantee at a deviation of at most 1
from the original value as the result of single bit flip error.}
\label{fig:intro:nn-code}
\vspace{-0.15in}
\end{figure}

\textit{Potential for end-to-end optimization.}
Once the network has been quantized and compressed, we can further
leverage resource versus correctness tradeoffs by storing the weights
in approximate SRAM~\cite{Reagen2016}, which occasionally produces
read errors. Recent work~\cite{Matic2018} shows that correct
re-training and fault detection mechanisms can mitigate the negative
effects of SRAM read upsets on classification tasks.

In addition, in the extreme case where weights are ternarized to
$-1$, $0$ and $1$, one could explore an encoding with a redundant
representation of zero as Figure~\ref{fig:intro:nn-code} shows.
With this encoding, single bit-flip errors would cause, in the worst
case, a deviation of one as opposed to a value polarity flip from
$-1$ to $1$ or vice-versa. The latter is allowed under the default
2's complement encoding, and could potentially lead to further
accuracy degradation.

\vspace{-0.05in}
\section{Terminology}
\label{section:terminology}
\vspace{-0.05in}
The terminology used to describe resource usage versus correctness
tradeoffs has historically differed across research communities
(e.g., the computer-aided design and design/test communities
versus the programming language and system software communities).
The differences in terminology are sometimes inevitable: a ``fault''
in hardware is usually a stuck-at logic- or device-level fault while
a ``fault'' in an operating system is usually the failure of a
larger macro-scale component. In this article, we attempt to provide
a uniform scaffolding for terminology. In doing so, we acknowledge
that this terminology will by necessity need to be reinterpreted
when applied to the different layers of abstraction in a computing
system and we do precisely that at the beginning of each of the 
following sections.

\newcommand{\domain}[1]{\mathsf{Domain}(#1)}
\newcommand{\range}[1]{\mathsf{Range}(#1)}

\vspace{-0.05in}
\subsection{Computation in the physical world}
\label{section:terminology:computation-and-physical-world}
\vspace{-0.05in}
%
%

\begin{figure}
\centering
\includegraphics[trim=0cm 0cm 0cm 0cm, clip=true, angle=0, width=0.85\textwidth]{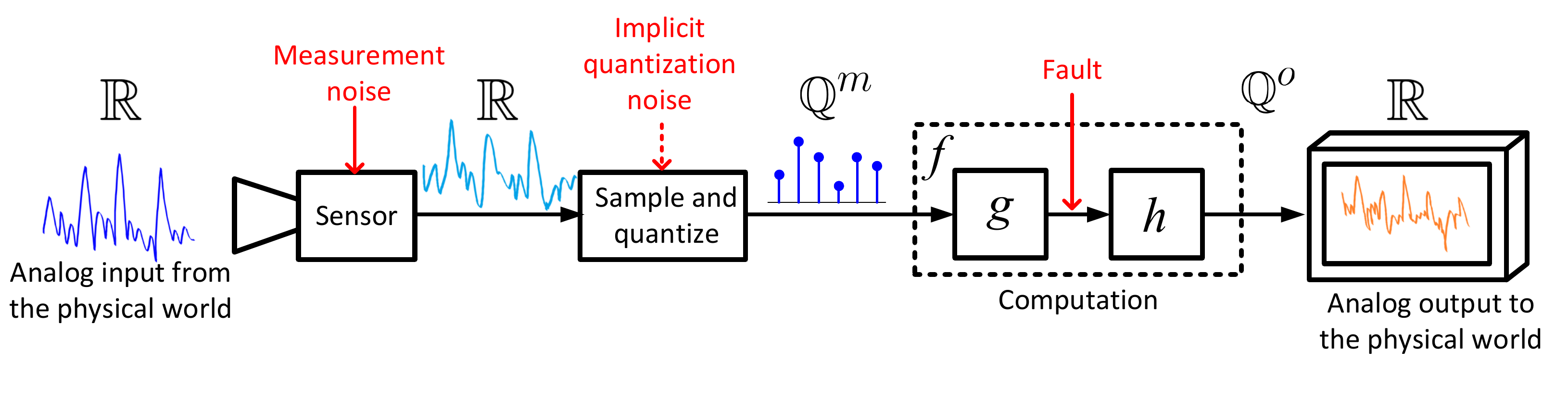}
\vspace{-0.25in}
\caption{Steps of computation in the physical world.}
\label{fig:analog_analog}
\vspace{-0.15in}
\end{figure}

We consider computing systems that make observations of the physical
world (e.g., using sensors or other data input sources) and compute
a discrete set of actions that the system (or a human) then applies
back to the physical world. Such end-to-end systems are therefore
\textit{analog in, analog out}.  In this process, the computing
system \textit{measures} the physical world, \textit{computes} on
a sample of its measurement, and then computes a set of \textit{actions
or actuations} to be applied to the world.  Figure~\ref{fig:analog_analog}
shows the steps of computation in the physical world.

%
%

We denote the domain of quantized values by $\mathbb{Q}$.  In
practice, quantized values are often bounded integers or finite-precision
floating-point numbers.  When working with a relation $r \subseteq
A \times B$, the domain and range of a relation $r$ are defined as
$\domain{r} := \{ x \,|\, \exists y. (x,y) \in f \}$ and $\range{r}
:= \{ y \,|\, \exists x. (x,y) \in f \}$.  The composition of two
relations $f$ and $g$, denoted $f\circ g$, is allowed if $\range{f}
\subseteq \domain{g}$ and it is defined as $f \circ g = \{ (x,z)
\,|\, \exists y. (x,y) \in f \land (y,z) \in g \}$.  A \emph{left-total}
relation is a relation that covers all members of its input domain.
In other words, an output exists for every possible input.

\textbf{Physical world:}\quad We assume that all of our systems are
situated in the physical world and we model inputs from this world
with real numbers, $\mathbb{R}$.  This assumption is consistent with
most applications that trade errors for efficiency (see
Section~\ref{section:applications} and Section~\ref{section:examples}),
such as sensing applications (as in Section~\ref{example:pedometer}),
cyber-physical systems, computer graphics, computer vision, machine
learning (as in Section~\ref{section:MNIST}), and scientific
computing.

\textbf{Measurement and analog processing step:}\quad Each computation situated in the physical
world begins with a \textit{measurement} in which the computing
system makes an observation of the physical world.  In metrology,
this quantity is referred to as the \textit{measurand}.
%
%
We denote the result of a measurement by a probability distribution.
We restrict our focus to distributions that we can represent with
a \textit{probability density function (PDF)}, $f : \mathbb{R} \rightarrow \mathbb{R}$.

Measurements may include within their internal processes computations
that transform the measured distributions to yield new distributions.
These internal processes may be nondeterministic. We include
this facility to account for systems that may perform computation
directly in the unsampled and unquantized analog domain and
Section~\ref{section:05:analog} of the survey gives examples of
such systems. A measurement is therefore a function of type $f :
\mathbb{R} \rightarrow (\mathbb{R} \rightarrow \mathbb{R})$, mapping
a real value (the measurand) to a function in the form of the
probability distribution (the measurement). The result of the
measurement step of a computation is therefore still in the domain
of continuous-time real-valued quantities.

\textbf{Sampling and quantization step:}\quad Between the measurement
step and a subsequent discrete (digital) computation step, we assume
that there is a sampling and quantization step that generates discrete-time samples with
discrete values from the real-valued distribution resulting from the
measurement step. 
A {\em sampler} is therefore a relation $f :
(\mathbb{R} \rightarrow \mathbb{R}) \times \mathbb{Q}^m$ that
samples and quantizes a discrete value from a probability distribution.
($\mathbb{Q}^m$ denotes the set of allowable quantized values.)
The process of quantization adds an implicit noise, known as the 
\emph{quantization noise} to the real-valued input.

\textbf{Digital computation step:}\quad In the discrete world, we consider the
computations that take as input a discrete sample from the measured world and
performs a potentially nondeterministic computation to produce a discrete
output.  Therefore, a discrete computation $f$ is a left-total relation $f
\subseteq \mathbb{Q}^m \times \mathbb{Q}^o$ where $\mathbb{Q}^m$ is the input
and $\mathbb{Q}^o$ is the output. 

\textbf{Actuation step:}\quad 
The digital outputs can be used back in the physical world as inputs
to real-valued actuation which modifies the state of the physical
world. An actuation computation is therefore a nondeterministic
function that we model as a left-total relation $f \subseteq
\mathbb{Q}^o \times \mathbb{R}$.

\vspace{-0.05in}
\subsection{Computation and correctness}
\vspace{-0.05in}
Following the terminology defined in
Section~\ref{section:terminology:computation-and-physical-world},
we can express any computation that processes data from the physical
world as a composition of the steps of measurement, sampling and
quantization, digital computation, and actuation. Each of these
steps defines a \textit{computation}:

\textbf{Computation:}\quad A computation $f$ is a nondeterministic function that we
denote as a left-total relation $f \subseteq \mathbb{I} \times \mathbb{Q}$ where we
instantiate the domain  $\mathbb{I}$ and $\mathbb{O}$ to fit the computation's
corresponding step from
Section~\ref{section:terminology:computation-and-physical-world}. For example,
as we will see later in
Section~\ref{section:transistorGateAndCircuitLevel:notation}, at the circuit
level, the input domain $\mathbb{I}$ and the output domain
$\mathbb{O}$ are voltage levels.

We model computations as left-total relations to account for nondeterminism
where for the same input, the computation may produce different outputs on
different executions. The relations are left-total in that there exists at
least one output for every value in the input domain. This modeling
assumptions also dictates that computations terminate.  If a computation is
deterministic, then we model it as a function $f : \mathbb{I} \rightarrow
\mathbb{O}$.

%
%

\textbf{Specification:}\quad For any computation $f \subseteq \mathbb{I}
\times \mathbb{O}$, a system's developers and users can provide its
\emph{specification} as a relation $f^* \subseteq \mathbb{I} \times \mathbb{O}$
that defines the set of {\em acceptable} mappings between the function's inputs
and outputs. A specification need not be executable itself and multiple
implementations can satisfy the same specification.

\textbf{Correctness:}\quad A computation and its corresponding definition as a
relation is \textit{correct} if it \textit{implements} its specification. A computation $f$
implements a specification $f^*$ iff  $ \forall i, o.\, (i,o) \in f \Rightarrow
(i,o) \in f^*$.  This definition means that every output of $f$ for a given
input, must be valid according to the specification.

\textbf{Faults:}\quad To define faults, we first decompose a computation $f$
into two computations $g \subseteq \mathbb{I} \times \mathbb{M}$ and $h
\subseteq \mathbb{M} \times \mathbb{O}$, where $\mathbb{M}$ is a domain of
values for the output of $g$ and $g \circ h \equiv f$. Given this
decomposition, a \textit{fault} is an anomaly in the execution of $g$ on an
input $i$ such that $g$ produces an anomalous, unexpected value $m$ in that 
$(i, m) \not\in g$.
 
\textbf{Errors:}\quad An error occurs when a computation encounters a fault and
the computation's resulting output does not satisfy its specification.  Given a
computation $f$ and its decomposition into $g$ and $h$ as above, the semantics
of an error is that if the execution of $g$ on $i$ produces a faulty value $m$
(as above), then that fault is an error if the result of the continued
execution via $h$ does not satisfy $f$'s specification\,---\,namely, that $(i,
h(m)) \not\in f^*$.

\textbf{Masking:}\quad A fault does not always result in an error; a fault
can instead be {\em masked}. If a computation encounters a fault and the
computation's resulting output satisfies its specification, then the fault has
been masked by the computation's natural behavior.  Given a computation $f$ and
its decomposition into $g$ and $h$ as before, the semantics of a masked error is that
if the execution of $g$ on $i$ produces a faulty value $m$ (as above), then
that fault is masked if the result of the continued execution via $h$ satisfies
$f$'s specification\,---\,namely, that $(i, h(m)) \in f^*$.

\textbf{Precision and accuracy:}\quad We define \textit{precision}
as the degree of discretization of the state space determined by
$\mathbb{Q}^m$ (from the sampling and quantization step,
Section~\ref{section:terminology:computation-and-physical-world})
and we define \textit{accuracy} as a distance between the functions
$f$ and $f^*$ defined above.

\vspace{-0.05in}
\subsection{Standard viewpoints}
\vspace{-0.05in}
Let $h$ be the identity function and $f=g$. Then the aforementioned
definitions give a semantics to faults that affect the output of a
single, monolithic function $f$. Take $f^*$ to be $f$, then the
function's specification is given by its exact behavior. This form
of specification is the standard assumption for computing systems
wherein they must preserve the exact semantics (up to nondeterminism)
of the computation.  Most existing approaches to trading errors for
efficiency fit this viewpoint: they typically start from an existing
program as their specification and approximate it to allow for more
efficient implementations.

\vspace{-0.05in}
\subsection{Quantifying errors}
\vspace{-0.05in}
Approaches to quantifying errors include absolute errors, relative
errors, and error distributions. In most contexts, the evaluator
of a system is interested in the error of not only a single input,
but a whole domain of inputs. Depending on the application domain,
upper or lower bounds on the worst-case error, or average errors
may be of interest.  When a computation runs repeatedly, the
\emph{error frequency} or \emph{error rate}
captures how often a computation returns an incorrect result.

\vspace{-0.05in}
\section{Transistor-, Gate-, and Circuit-Level Techniques}
\label{section:transistorGateAndCircuitLevel}
\vspace{-0.05in}

%
%
Transistors provide the hardware foundations for all computer
systems. As a result, their physical properties determine the
efficiency at which a particular computation can be performed.  When
collections of transistors are used to form gates and analog circuits,
and when collections of gates are used to implement digital logic
circuits, the organization of the transistors, gates, and circuits
can be designed to trade efficiency for correctness.

%
%
%

\vspace{-0.05in}
\subsection{Notation}
\label{section:transistorGateAndCircuitLevel:notation}
\vspace{-0.05in}
Following the notation introduced in Section~\ref{section:terminology},
input $\mathbb{I}$ can be defined as a voltage level that is switched
to $\mathbb{O}$ as a computation $f$ is executed. Therefore, $f:
\mathbb{I} \rightarrow \mathbb{O}$ is a switching of voltage
at a transistor or a group of transistors forming a circuit element,
for instance, a byte in a memory or an adder in an arithmetic/logic unit (ALU).
Such computation can be regarded as $f^*:
\mathbb{I} \rightarrow \mathbb{O}$ where the relation between
$f^*$ and $f$ is the difference in the electrical operating points
of the individual transistors. This difference saves
computational costs like power consumption and latency while
introducing timing errors and incorrect voltage levels.

\vspace{-0.05in}
\subsection{Analog input / analog output systems: a comparison reference for quantization}
\vspace{-0.05in}
When using finite-precision arithmetic, computation always involves
errors that are caused by quantization. Quantization is a fundamental
mechanism for trading energy for accuracy and recent work has highlighted
examples of its effectiveness~\cite{MoreauQuantization}.

The effect of quantization errors can be observed by treating the
inputs and outputs of a computing system as real-valued analog
signals and comparing these signals to an ideal (error-free) computing
system that accepts analog inputs and produces analog outputs.  When
such ideal outputs are not available, designers often use the output
of the highest precision available (e.g., double-precision floating
point) as the reference from which to determine the error of a
reduced-precision block.  Such analyses are common in the design
process of digital signal processing algorithms such as
filters~\cite{ProakisDSP} where the choice of number representation
and quantization level enables a tradeoff between the performance
and signal-to-noise ratio properties of a system.

\vspace{-0.05in}
\subsection{Analog computing: data processing with currents and voltages}
\label{section:05:analog}
\vspace{-0.05in}
Analog computing systems~\cite{cct:MacLennan, karplus1959analog} eliminate the need for discretization and the
resulting restriction on precision that is inherent in digital
circuits.  While, in theory, analog circuits provide unbounded
precision, in practice their precision is limited by factors such
as noise, non-linearities, the degree of control of properties of
circuit elements such as resistors and capacitors, and the degree of control of
implicit parameters such as temperature.  At higher precision,
analog blocks tend to be less energy-efficient than digital blocks
of equivalent precision~\cite{sarpeshkar1998analog}. Because they
usually do not use minimum-size transistors, analog circuits may
also be larger in area than their digital circuit equivalents.
Designing analog computation units is also a challenging task.
Nevertheless, analog circuits can be an attractive solution for 
applications that tolerate low-precision computation~\cite{sarpeshkar1998analog}.

\vspace{-0.05in}
\subsection{Probabilistic computing: exploiting device-level noise for efficiency}
\vspace{-0.05in}
A line of research pioneered by Palem\etalii~\cite{cct:p1, cct:p2}
(``probabilistic computing'') proposes harnessing intrinsic thermal
noise of CMOS circuits to improve the performance of probabilistic
algorithms that exploit a source of entropy for their execution.
Chakrapani\etalii~\cite{cct:p3} show an improvement in the energy-performance
product for algorithms such as Bayesian inference, probabilistic
cellular automata, and random neural networks using this
approach and they establish a tradeoff between the
energy consumption and the probability of correctness of a circuit's
behavior. These techniques have also shown energy savings for digital
signal processing algorithms that do not employ probabilistic
algorithms but which can tolerate some amount of noise~\cite{cct:p4,
cct:p5}.

\vspace{-0.05in}
\subsection{Stochastic computing: unary representation and computing on probabilities}
\vspace{-0.05in}
Stochastic computing (SC) uses a data representation of
bit streams that denote real-valued probabilities~\cite{AlaghiTCAD17}.
In theory, the probabilities can have unbounded precision, but in
practice, the length of the
bit-streams determines precision~\cite{cct:SC}. SC was first
introduced in the 1960s~\cite{cct:Gaines67} and its main benefit is
that it allows arithmetic operations to be implemented via simple
logic gates: a single {\sc and} gate performs SC multiplication.
This made SC attractive in the era of expensive transistors. But
as transistors became cheaper, SC's benefit faded away, and its
main drawbacks, i.e., limited speed and precision, became
dominant~\cite{cct:SC}. For this reason, SC was only used in certain
applications, such as neural networks~\cite{Dickson93, Kim95} and
control systems~\cite{Toral00}.

SC has seen renewed interest over the last decade~\cite{cct:SC},
mainly because of its energy efficiency. SC's probabilistic nature
copes with new inherently random technologies such as
memristors~\cite{Knag14}. Furthermore, the unary encoding of numbers
on SC makes the computation robust against errors~\cite{Qian11},
and allows variable precision computation~\cite{Alaghi14}. With the
low precision requirement of modern machine learning applications,
SC is becoming an attractive alternative to conventional binary-encoded
computation~\cite{cct:SCNN}.

%
%
%

%
%
\setlength{\abovedisplayskip}{3pt}
\setlength{\belowdisplayskip}{3pt}

\vspace{-0.05in}
\subsection{Voltage overscaling: improved efficiency from reduced noise margins}
\label{section:05:overscaling}
\vspace{-0.05in}
The term \emph{voltage overscaling} is often used to refer to reducing
supply voltages more than is typically deemed safe for a given clock
frequency.  Voltage overscaling exploits the quadratic relationship
between supply voltages and dynamic power dissipation. Let
$V_{\mathrm{dd}}$ be the supply voltage of a CMOS circuit (e.g.,
an inverter), let $f$ be its clock frequency (reciprocal of its
delay) and let $C$ be the effective capacitance of the load of the
circuit. Then, the dynamic power dissipation $P$ is~\cite{Rabaey96}
\begin{align}
	P \propto C V_{\mathrm{dd}}^2 f .
\end{align}
The delay of a gate in a circuit, and hence the clock frequency
$f$, is however not independent of supply voltage $V_{\mathrm{dd}}$.
Let $V_t$ be the device threshold voltage and let $\alpha$ be a
process-dependent parameter (the velocity saturation
exponent~\cite{Sakurai:alpha}). Then, as supply voltage $V_{\mathrm{dd}}$
decreases, the delay of charging its load capacitance for a gate
increases and the maximum clock frequency achievable at a given
voltage follows the relation
\begin{align}
	f \propto \frac{(V_{\mathrm{dd}} - V_t)^\alpha}{V_{\mathrm{dd}}} .
\end{align}
As a result, overscaled voltages cause circuit delays, which in
turn lead to timing errors in circuits at a fixed clock speed.
Several approaches have explored the idea of carefully and
systematically accepting such errors in exchange for the large
(quadratic) power savings that voltage overscaling can potentially
provide~\cite{cct:shanhbag:sdsp, cct:shanhbag:sedsp, cct:kurdahi:vos,
cct:roy:vos, cct:Kahng:vos}.  In unmodified circuits, this often
leads to catastrophic errors at close-to-nominal voltages, as many
digital circuits are optimized to minimize timing slack. However,
for several application domains, such as image and video processing,
inherent dependence of errors on known input characteristics can
be exploited to redesign circuits such that they allow for significant
overscaling with small and graceful degradation of output
quality~\cite{cct:shanhbag:motion, cct:roy:motion, cct:roy:dct,
cct:TERRA}. However, voltage overscaling has potential issues with timing closure and meta-stability. Furthermore, timing errors
in the critical paths of a circuit due to voltage overscaling tend to affect the most significant bits
of a computation first and hence can lead to large errors. 
Dedicated logic modifications targeting lower significant bits as described next can instead provide
better accuracy with additional switching activity savings for the
same timing and hence voltage reduction~\cite{cct:Dith}.

\vspace{-0.15in}
\subsection{Pruned circuits for efficiency at the expense of precision and accuracy}
\vspace{-0.05in}
Pruning circuits refers to deleting or simplifying parts of a circuit
based on the probability of their usage or importance to
output quality. Recent research has shown how circuit pruning
improves latency, energy, and area without the overheads
associated with voltage scaling~\cite{cct:p6, cct:p7, cct:ls1,
cct:ALS}.

Pruning can be applied to digital circuit building blocks such as
adders and multipliers, enabling quality-cost tradeoff opportunities
through different logic simplification and pruning techniques.
Approximate adders attempt to simplify carry chains~\cite{cct:ETA,
cct:Specul} or to use approximate 1-bit full adders~\cite{cct:AMA,
cct:LOA, cct:Dith} at lower significant digits of a sum computation.
Accuracy-configurable adders have also been proposed for adaptive-accuracy
systems that require a functional unit like an adder or multiplier
to vary the degree of tradeoff between correctness and resource
usage based on the quality demand of computation~\cite{cct:config1}.
Unlike approximate adders, approximate multipliers have a higher
design space exploration requirement, as they are composed of
2$\times$2 partial products that are summed up by deploying an adder
tree to compute the final result~\cite{cct:Mul1}.  Correctness
versus resource usage tradeoffs can be deployed in multipliers
(partial products) or adders, or both, for a chosen number of
least-significant bits~\cite{cct:Mul2, cct:Mul3}.


Approximate adders and multipliers provide the combinational building blocks for
approximate datapath and processor designs. At the sequential logic
level, the challenge is in determining the amount of approximation
to apply to each addition or multiplication operation in a larger
computation in order to minimize output quality loss while maximizing
energy savings. For example, in a larger computation that consists
of multiple accumulations, using an adder with a zero-centered error
distribution~\cite{cct:Dith} will result in positive and negative
errors canceling each other and thus averaging in the final output
of a larger accumulation. By contrast, in other computations, an
approximate adder that always over- or under-estimates may be
beneficial.

Determining the best tradeoff for each functional unit has been
investigated for fixed register transfer level (RTL)
designs~\cite{cct:ABACUS, cct:SALSA}.  Pure RTL optimizations, on
the other hand, do not exploit changes in approximated component
characteristics for a complete RTL re-design.  In the context of
custom hardware/accelerator designs, selection of optimal approximated
operator implementations can instead be folded into existing C-to-RTL
high-level synthesis (HLS) tools~\cite{cct:AHLS, cct:AHLS2}. For
programmable processors, accuracy configuration of the datapath can
be exposed through the instruction-set architecture (ISA)~\cite{cct:Quora}.
A compiler then has to determine the precision of each operation
in a given application (see
Section~\ref{section:architectureAndMicroarchitectureLevel} and
Section~\ref{section:programmingLanguage}).

\vspace{-0.05in}
\subsection{Approximate memory: reducing noise margins for efficiency in storage}
\label{section:devices:approximateMemory}
\vspace{-0.05in}
Memory costs are often higher than that of computations in many
data-intensive applications~\cite{EnergyProportional}.  Approximate
memories have been investigated in the research literature,
to trade quality for energy, latency, and lifetime
benefits~\cite{cct:cache, sampson:approximate}.  Reducing the refresh
rate of DRAM provides an opportunity to improve energy
efficiency while causing a tolerable loss of quality~\cite{cct:DRAM}
(Section \ref{section:architecture:approximateMemory}).  For static
random access memory (SRAM) on the other hand, the tradeoff between
correctness and resource usage is typically achieved by voltage
overscaling, where the main concern is in dealing with the failures
in the standard 6-Transistor (6T) cells of an SRAM array under
reduced static noise margins (SNMs)~\cite{cct:SRAM}.  As a result,
hybrid implementations combining 6T with 8T SRAM cells~\cite{cct:hybrid1}
or with standard cell memory (SCMEM)~\cite{cct:hybrid2} have been
employed to achieve aggressive voltage scaling in order to get
better quality versus cost tradeoffs.

\vspace{-0.05in}
\subsection{Summary}
\vspace{-0.05in}
%
%
The circuit-level techniques surveyed in this section must ultimately
be deployed in the context of concrete applications. For example,
one case study found that for applications such as Fast Fourier
Transforms (FFTs), motion compensation filters, and $k$-means clustering,
applying traditional fixed-point optimizations to limit the size
of operands was more effective than applying circuit-level
approximations such as approximate adders and multiplier
circuits~\cite{cct:c1}.  This is because approximating some bit
values still requires information about those bits to be stored and
used in downstream computations. The additional overhead of this
bookkeeping in many cases is not worth the quality benefits.
Carefully selecting the most suitable approximation strategies and
comparing their cost versus quality tradeoffs can therefore lead to a
better solution for certain applications.



\vspace{-0.05in}
\section{Architecture and Microarchitecture-Level Techniques}
\label{section:architectureAndMicroarchitectureLevel}
\vspace{-0.05in}
Architectural and microarchitectural techniques that
trade correctness for resource usage have focused primarily on
correctness at the software or application level and
have focused on reducing resource usage in memory, in the processor, and in on- or off-chip I/O.

%
%
%

\vspace{-0.05in}
\subsection{Notation}
\vspace{-0.05in}
Architectural techniques create abstractions that allow operating
systems, programming languages, and applications to specify their
precision and accuracy requirements through specialized instructions
and instruction extensions. Following the notation introduced in
Section~\ref{section:terminology}, the computation function $f :
\mathbb{Q}^m \rightarrow \mathbb{Q}^o$ is defined over the quantized 
sets $\mathbb{Q}^m$ and $\mathbb{Q}^o$
embodied by software-visible machine state such as registers, memory,
and storage. The computation function $f$ is implemented using
either general-purpose cores or specialized hardware accelerators.
Microarchitectural techniques facilitate the efficient implementation
of the computation function $f$ at the level of hardware functional
units, such as memory controllers and processor pipelines, or by
the efficient hardware representation of the sets $\mathbb{Q}^m$ and $\mathbb{Q}^o$.

\vspace{-0.05in}
\subsection{Trading resource usage for correctness in processor cores}
\vspace{-0.05in}
Early work trading resource usage for correctness such as Razor and
related techniques~\cite{ernst:razor, sridharan:eliminating}, relied
on voltage overscaling as the primary underlying circuit-level
mechanism to increase energy efficiency. As a
result, these techniques provided no direct means to improve
performance, but provided higher energy efficiency at the expense
of nondeterministic faults. To mask such faults and hide them from
applications, voltage overscaling approaches typically rely on
error recovery mechanisms. The key insight is that sophisticated
error recovery mechanisms can be much more resource-efficient in
ensuring correctness compared to voltage over-provisioning. Carefully
balancing the error recovery overhead against the benefits of voltage
overscaling can provide higher energy efficiency without sacrificing
output quality or program safety~\cite{ernst:razor,
sridharan:eliminating}.

Truffle~\cite{esmaeilzadeh:architecture} was the first architecture
to willingly introduce uncorrected nondeterministic errors in
processor design for the sake of energy efficiency.  Truffle uses
voltage overscaling selectively to implement approximate operations
and approximate storage. The Truffle architecture provides ISA
extensions to allow the compiler to specify approximate code and
data and its microarchitecture provides the implementation of
approximate operations and storage through dual-voltage operation.
For error-free operations, a high voltage is used, while a low
voltage can be used for approximate operations. Voltage selection
is determined by the instructions, with the control-flow and address
generation logic always operating at a high voltage to ensure safety.

In addition to improving energy efficiency, architectures that
enable tradeoffs between resource usage and correctness may result
in higher performance compared to an error-free baseline~\cite{amant:general,
esmaeilzadeh:neural, yazdanbakhsh:neural, moreau:snnap}.  Examples
of approaches include offloading parts of a processor's workload
to computing units that can perform the desired functionality much
faster at the cost of deviation from correct behavior. Because of
their performance advantage, such computing units are often called
accelerators.  Accelerators that trade resource usage for correctness
include, most notably, neural accelerators~\cite{amant:general,
esmaeilzadeh:neural, yazdanbakhsh:neural, moreau:snnap}, which
implement a hardware neural network trained to mimic the output of
a desired region of code.

Temam\etalii empirically show that the conceptual error tolerance
of neural networks translates into the defect tolerance of hardware
neural networks~\cite{temam:defect}, paving the way for their
introduction in heterogeneous processors as intrinsically error-tolerant
and energy-efficient accelerators. St. Amant\etalii  demonstrate a
complete system and toolchain, from circuits to a compiler, that
features an area- and energy-efficient analog implementation of a
neural accelerator that can be configured to approximate general
purpose code~\cite{amant:general}. The solution of St. Amant\etalii
comes with a compiler workflow that configures the neural network's
topology and weights.  A similar solution was demonstrated with
digital neural processing units, tightly coupled to the processor
pipeline~\cite{esmaeilzadeh:neural}, delivering low-power approximate
results for small regions of general-purpose code. Neural accelerators
have also been developed for GPUs~\cite{yazdanbakhsh:neural}, as
well as FPGAs~\cite{moreau:snnap}.

\vspace{-0.05in}
\subsection{Approximate memory elements}
\label{section:architecture:approximateMemory}
\vspace{-0.05in}
Memory architectures that trade resource usage for correctness
permit the value that is read from a given memory address to differ
from the most recent value that was written. The traditional view
of memory elements assumes that every memory access pair consisting
of a write followed by a subsequent read operation, applied to a
input $\mathbb{I}$, results in the same read result for a given
write value.  In contrast, approximate memory
elements may perform non-identity transformations of the input
$\mathbb{I}$. The benefits of doing so include reduced read/write
latency, reduced read/write access energy, fewer accesses to memory,
increased read/write bandwidth, increased
capacity~\cite{sampson:approximate, guo:high, jevdjic:approximate},
improved endurance~\cite{sampson:approximate}, and reduced leakage
power dissipation~\cite{liu:flikker}. These techniques have been
applied to memory components ranging from CPU
registers~\cite{esmaeilzadeh:architecture},
caches~\cite{esmaeilzadeh:architecture, sanmiguel:load,
sanmiguel:doppelganger, sanmiguel:bunker}, main memory~\cite{liu:flikker},
to flash storage~\cite{sampson:approximate, guo:high, jevdjic:approximate}.

One method for trading resource usage for correctness in memories
is to predict memory values instead of performing an actual read
operation.  For example, on the occurrence of a cache miss,
\textit{load value approximation} (LVA)~\cite{sanmiguel:load,
thwaites:rollback} provides predicted data values to a processor
which may differ from the correct values in main memory. Doing so 
hides cache miss latency and thereby reduces the average memory
access time at the expense of having data values in the cache that
differ from what they would be had they been faithfully loaded from
main memory. The correct values in main memory may subsequently be
read from memory to train the predictor and improve its accuracy,
or the main memory access may be skipped entirely to save energy.
Conventional value prediction considers any
execution relying on predicted values speculative and provides
expensive microarchitectural machinery to roll back execution in the case of
a mismatch between the predicted and actual values. LVA, by contrast, allows
imperfect predictions, trading correctness of values in the cache
for reduced micro-architectural complexity and reduced memory
latency.

Several memory technologies expose circuit-level mechanisms to trade
accuracy for reduction in latency or access energy (or both). For
example, multi-level solid-state memories perform write operations
iteratively, until the written value is in the desired range. By
reducing the number of write iterations, approximate
writes~\cite{sampson:approximate} significantly reduce the latency
and energy of write operations, increasing write bandwidth as a
side effect, at the expense of reduced data retention. In spintronic
memories, such as STT-MRAM, reducing the read current magnitude can
reduce energy of read operations at the expense of accuracy of the
content being read.  In contrast, significantly increasing the read
current magnitude reduces the read pulse duration, decreasing the
read latency while potentially disturbing the written content with
noise. Such mechanisms can be leveraged at the architectural level
through dedicated instructions for imprecise loads and
stores~\cite{ranjan:approximate}.

The correctness of values obtained from memories can also be traded
for an increase in effective storage capacity. One way to achieve this is
to avoid storing similar data multiple times. For example, storing
similar data in the same cache line can save on cache space in
situations when substituting a data item for a similar one still
yields acceptable application quality~\cite{sanmiguel:doppelganger,
sanmiguel:bunker}.  Another way to trade errors for capacity is
through deliberate reduction in storage resources dedicated to
error-correction~\cite{guo:high, jevdjic:approximate}. By providing
weaker error-correction schemes for data whose accuracy does not
have a critical impact on the output quality, significant storage
savings have been demonstrated in the case of encoded images and
videos~\cite{guo:high, jevdjic:approximate}.

For volatile memory technologies, such as SRAM and DRAM, voltage
scaling approaches can be used to reduce the static energy at the
expense of faults, observed as bit flips~\cite{esmaeilzadeh:architecture}.
In the case of DRAM, the energy needed to retain data can be further
reduced by less frequently refreshing the DRAM rows that contain
data whose incorrectness applications can tolerate, compared
to the rows that applications require to remain correct~\cite{liu:flikker}. 
In solid-state memories, mapping data that applications can tolerate to
be incorrect onto blocks that have exhausted their hardware error 
correction resources can significantly extend endurance~\cite{sampson:approximate}.

\vspace{-0.05in}
\subsection{Approximate communication}
\vspace{-0.05in}
As in the case of approximate memory elements, approximate communication
systems may perform non-identity transformation $f^*$ of input
$\mathbb{I}$ to efficiently transfer the input through a communication
channel or network. The idealized computation function $f$ corresponds
to an identity transformation over an infinitely large input.
Examples of inputs include signals on intra- and inter-chip wires,
such as memory buses and on-chip networks. The architectural
techniques trading resource usage for correctness in such systems
usually rely on more efficient but less reliable links, network
buffers, and other network elements, or employ lossy in-network
compression to minimize data movement, while overlapping the
compression and communication.

The conventional approach to trade resource usage for correctness
in communication over a channel is to employ lossy compression at
the source and decompression at the destination, with the goal of
reducing the amount of data transferred through the channel, as
well as to reduce latency. Such approaches have been
widely used for decades in long-distance communication, such as
media streaming applications. However, when the communicating parties
are two processors on a board, two cores on a chip, or a core and
a cache, the communication latency is in the order of nanoseconds
and any compression/decompression latency added to the critical
path of program execution may be prohibitive.

At the circuit level, transmitting bits over a wire on-chip or over
a printed circuit board trace costs energy. For single-ended I/O
interfaces, where the signaling of information is with voltage
levels, the energy cost is typically due to the need to charge the
wire capacitance when driving a logic '1', and to discharge that
capacitance when driving a logic '0'. Building on this observation,
and on the body of work on low-power bus
encodings~\cite{Stan:1995,cheng2001memory}, value-deviation-bounded
serial encoding (VDBS encoding)~\cite{stanley2015efficiency,
Stanley-Marbell:2016:RSI:2897937.2898079} trades correctness for
improved communication energy efficiency by lossy filtering of
values to be transmitted on an I/O link. VDBS encoding reduces the number of '0'
to '1' and '1' to '0' signal transitions and hence reduces the
energy cost of I/O.  Because VDBS encoding requires no decoder, it
can be implemented with low overhead, requiring less than a thousand
gates for a typical implementation~\cite{hotchips16encoder}.
Extensions of VDBS encoding have extended the basic concept to
exploit temporal information in information streams~\cite{kim2017axserbus,
pagliari2016serial} and to employ probabilistic encoding
techniques~\cite{1804.02317}.

A recent study leverages data similarity between cache blocks to
perform lossy compression in networks-on-chip
(NoCs)~\cite{boyapati:approxnoc}.  The key idea is in simple data-type
aware approximation using approximate matching between data to be
sent and data items that have been recently sent to perform a quick
lossy compression.  Performing approximation at the network layer
allows a significant data movement reduction without losing the
precise copy of the data and without extending the critical path,
as the communication and compression are overlapped.

An orthogonal approach to trading resource usage for correctness
in communication by compression, is to reduce the safety margins
of communication links to trade off their reliability for bandwidth,
latency, or both. For on-chip networks, achieving reliable transmission
in low-latency high-bandwidth interconnects requires features like
forward error correction (FEC), but FEC can increase communication
latency, by up to three fold in one study~\cite{fujiki:highbandwidth}.
An approach to counteract such high overheads is to allow higher
bit error rates at the link layer by removing forward error correction
or employing a weaker but more efficient error correction mechanisms,
with a variable amount of redundancy based on application
needs~\cite{fujiki:highbandwidth}.  A low-diameter network is one
approach to keep the end-to-end bit error rate under control,
minimizing the number of hop counts, and thus prevent excessive
accumulation of errors~\cite{fujiki:highbandwidth}.

Allowing errors in communication can be particularly challenging
in parallel programs, which rely on communication for synchronization.
In such contexts, failure to deliver correct messages on time can
affect control flow and lead to catastrophic
results~\cite{yetim:commguard}. Yetim\etalii propose a mechanism
to mitigate inter-processor communication errors in parallel programs
by converting potentially catastrophic control flow or communication
errors into likely tolerable data errors~\cite{yetim:commguard}.
Their main insight is that data errors have much less impact on the
application output compared to errors in control flow. Their approach
is to monitor inter-processor communication in terms of message
count, and to ensure that the number of communicated items is correct,
either by dropping excess packets or by generating additional packets
with synthetic values. Ensuring the correct number of exchanged
messages improves the integrity of control flow
in the presence of communication errors and consequently improves
the output quality of approximate parallel programs.

%
%

\vspace{-0.05in}
\subsection{Summary}
\vspace{-0.05in}
Microarchitectural techniques that trade correctness for resources
build on circuit level techniques
(Section~\ref{section:transistorGateAndCircuitLevel}) to exploit
information at the level of hardware structures such as caches,
register files, off-chip memories, and so on. Architectural techniques
expose microarchitectural techniques to software through constructs
such as instruction extensions, new instruction types, or new
hardware interfaces to accelerators. Exposing information about
hardware techniques to software allows software to take advantage
of the implemented techniques, while exposing information from
software to hardware allows hardware to, for example, more aggressively leverage 
tradeoffs between correctness and resource usage. In the
same way that circuit-level techniques form a foundation for the
approaches discussed in this section, circuit-level, microarchitectural,
and architectural techniques similarly form a foundation for
operating system and runtime system techniques.

\vspace{-0.05in}
\section{Programming Language Techniques}
\label{section:programmingLanguage}
\vspace{-0.05in}
Many programming-language- and compiler-level techniques that
trade correctness for efficiency provide abstractions for dealing
with errors introduced at lower levels of the system stack, or
introduce higher-level approximations directly and these errors
combine into whole-application errors. For these whole-application
errors, another set of techniques provide methods for reasoning
about errors and for managing them.

\vspace{-0.05in}
\subsection{Notation}
\vspace{-0.05in}

Following the notation introduced in Section~\ref{section:terminology},
programming language techniques usually operate at a level of
abstraction where the computation's implementation $f : \mathbb{I} \times \mathbb{O}$
is defined over a sets represented by, e.g., integers or
floating-point numbers. These integer and floating-point number
representations serve as an abstraction for the actual bit-level
representations of program state in hardware.  The idealized specification
 $f^*$ that $f$ implements may thus, for instance, assume unbounded
integers or real numbers for its output $\mathbb{O}$ and may represent an
entire computational problem or specific algorithm.  Examples of
errors introduced by the discrepancy between $f$ and $f^*$ are
floating-point roundoff errors, errors due to skipping entire
portions of a computation or due to missing synchronization.  The
appropriate means of measuring error, e.g., absolute error, relative
error, worst-case error, or error probability, is typically
application-dependent.

\vspace{-0.05in}
\subsection{Static compile-time techniques}
\vspace{-0.05in}
Static techniques aim to make resource versus correctness tradeoffs
safe to apply without having to run a program. To achieve this,
they isolate the effects of errors, or quantify the magnitude or
probability of errors at compile time.
Errors introduced at the lower levels of the stack do not affect
every operation of a high-level program equally. Ideally, errors
in lower layers of the stack should be restricted locations such
that when they propagate to higher layers of the stack, they only
affect those parts of the program where errors can be
tolerated.

Traditional programs, however, do not provide a transparent way to
mark what can be potentially approximate.  EnerJ~\cite{Sampson2011}
and FlexJava~\cite{Park2015} make the effects of lower-level errors
explicit by allowing programmers to annotate values in programs
that can be potentially affected by errors. They then use type
inference and taint analysis, respectively, to model propagation
of errors through a program automatically, minimizing the need for
programmers to explicitly trace through their programs to identify
all locations where errors could have an effect.  In a similar vein,
the Uncertain<T> type~\cite{Bornholt2014} encapsulates probability
distributions, e.g., resulting from measurement errors from a sensor.
The Uncertain<T> type system only allows a small number of specifically-designed
operations on values tagged with this type, allowing programmers
to be aware of which variables in their programs take on distributions
of values which are uncertain.

Rely~\cite{Carbin2013} provides a sound probabilistic reasoning
framework, i.e., a set of rules which allow a programmer to derive
the probability of a result being exact, given the probabilities
of individual operations being exact. Rely's reasoning framework
is guaranteed to be correct. Boston\etalii~\cite{Boston2015} provide
an automated system to obtain the probabilities required by Rely,
by encoding the task of determining the probabilities as a
type-inference problem.

The probability of a computed value being incorrect does not capture
the numeric magnitude of the error. Numeric error estimation has been
addressed in the form of static analysis for bounding errors due
to input disturbances~\cite{Chaudhuri2011} and optimizing
finite-precision arithmetic while guaranteeing sound error
bounds~\cite{Chiang2017,Darulova2018}.  Numeric error magnitude can
also be estimated by differential program verifiers to check relative
safety, accuracy, or termination with respect to some reference
implementation by reduction to a \emph{satisfiability modulo theories}
(SMT) problem~\cite{He2016}.

The above approaches either quantify the probability or the error
magnitude, but not both. Furthermore, they do not optimize directly
for performance or energy usage.  Chisel~\cite{Misailovic2014}
combines a reliability analysis with error bounds computation. It
automatically finds approximations satisfying a specification and
minimizes energy by reduction to an ILP problem. Zhu\etalii~\cite{Zhu2012} 
propose a framework which explores a randomized combination of 
resource-correctness tradeoffs provided by a user.  It presents a 
tradeoff space exploration algorithm based on linear programming, 
which provides near-optimal guarantees.

Static techniques are desirable as they can provide strong correctness
guarantees. However, for a static optimization technique, a faithful
resource cost model is needed. Until now, these models have been
mostly high-level, coarse, and additionally not consistent across
different techniques or evaluations, making combinations and
comparisons of different techniques challenging.  These models
necessarily have to abstract over the underlying hardware in order
to be scalable and widely applicable, but they also need to reflect
reality as much as possible. Here, a tighter collaboration between
the software and hardware is needed (see Challenge 2 in
Section~\ref{section:challenges}).

\vspace{-0.05in}
\subsection{Dynamic runtime techniques}
\vspace{-0.05in}
Static guarantees are in practice achievable only for small programs.
For many applications such strong guarantees may not be necessary.
Dynamic or testing-based validation techniques trade correctness
for practical scalability and have been widely used to identify
resource versus correctness tradeoffs and to validate the quality
of these tradeoffs.

A first step when implementing an application in an error-efficient
way is to determine which parts of the application are resilient
to errors and which are not~\cite{rinard2006,rinard2007}. Different
applications allow for error-efficient computing to various degrees.
For instance, some algorithms can tolerate higher error rates but
lower error magnitudes and vice versa~\cite{chippa2013analysis}.

Profiling has traditionally been used to identify performance-intensive
portions of a program. A quality of service profiler~\cite{Misailovic2010}
takes into account quality of the results in addition to performance
and can thus identify resilient portions of an application.  A
similar idea has been explored by the Application Resilience
Characterization (ARC) framework~\cite{chippa2013analysis}, which
profiles an application while injecting errors, derives a
statistical error-resilience profile, and identifies the best
error-efficient technique for the given application.  The statistical
error-resilience profile has also been explored for iterative
workloads~\citep{gillani2017improving} to identify the number of
resilient iterations.

Once resiliency of an application is established, different techniques
can be applied which trade resource usage for correctness.  For
instance in arithmetic, Precimonious~\cite{Rubio-Gonzalez2013}
assigns differing floating-point precisions across the variables
in a program.  STOKE~\cite{Schkufza2014}, on the other hand, generates
entirely new implementations of floating-point functions. Both
Precimonious and STOKE ensure that on a given test set a user-defined
quality bound is satisfied.  Building on the observation that loops
usually make up the bulk of running time of a program, loop
perforation~\cite{Misailovic2010, Sidiroglou-Douskos2011} selectively
skips entire loop iterations.  Synchronization is another expensive
part of many applications, and several research efforts have observed
that some synchronization primitives can be removed without impacting
quality significantly~\cite{Renganarayana2012, Misailovic2012}.
Misailovic\etalii~\cite{Misailovic2013} explore nondeterminism as
a technique for trading resource usage for correctness techniques,
by parallelizing a sequential program such that data races can
occasionally occur.

Another approach to exploiting resilient applications is to let a
user define several application components with different
resource-correctness tradeoffs and to provide tool support to select
between these candidates to obtain a final implementation~\cite{Ansel2009,
Ansel2011, Fang2014}.   Neural networks can also be used to
replace blocks of imperative code~\cite{esmaeilzadeh:architecture}
and can provide a performance benefit when coupled with a dedicated
neural processing unit.  The Intel open-source approximate computing
toolkit (iACT)~\cite{mishra2014iact} provides a simulation-based
testbed for different approximations, such as precision scaling and
approximate memoization.

Although not all resource versus correctness tradeoffs are suitable
for all application domains, most of the techniques discussed above
are application-independent. Chippa\etalii~\cite{Chippa2010} and
Venkataramani\etalii~\cite{Venkataramani2015} present application-specific
approaches for machine learning classifiers which exploit the fact
that many instances are easy to classify. These easy-to-classify
instances are handled by simpler classifiers, while harder-to-classify
instances use increasingly more complex classifiers.
SAGE~\cite{Samadi:2013:SSA:2540708.2540711} and Paraprox~\cite{Samadi2014}
are specific to data-parallel kernels running on GPUs and provide
specialized approximate versions of common idioms, such as maps and
reductions.

\vspace{-0.05in}
\subsection{Summary}
\vspace{-0.05in}
The techniques discussed above are first steps towards addressing
the need for automated tool support for developers (Challenge 3 in
Section~\ref{section:challenges}) but they remain limited because
each addresses one particular point in the design space. More
comprehensive tools and ways to combine the existing techniques are
necessary. One solution might be for researchers to make their
program analyses and program transformations available as passes
for the LLVM compiler infrastructure.

Today, many techniques employ a simplified model of the underlying
hardware and these models are rarely based on characterization of
real hardware systems. In the future, error models will need to be
consistent with the errors observed at the hardware level. In
addition to these extensions of the way software-level techniques
are evaluated today, end-to-end evaluation platforms could provide
increased confidence in research results (Challenges 2 and 8 in
Section~\ref{section:challenges}).

%
%

\vspace{-0.05in}
\section{Operating System and Runtime Techniques}
\label{section:OsLevel}
\vspace{-0.05in}
Operating system (OS) and runtime techniques for trading correctness
for efficiency dynamically monitor a running system and adapt its
accuracy to a changing environment. These systems may take explicit
input from a program, such as through an application programming
interface (API) or system call interface, or might be driven based
on user input.

\vspace{-0.05in}
\subsection{Notation}
\vspace{-0.05in}
For OS and runtime techniques, measuring, sampling, and quantizing
signals from the physical world are already completed by the lower
layers of the system.  Following Section~\ref{section:terminology},
computation is a nondeterministic function $f^* : \mathbb{Q}^m
\rightarrow \mathbb{Q}^o$ with nondeterminism introduced by the
need to multiplex processes over a shared resource (the processor)
in the presence of asynchronous input and output events, user
interaction, and time-varying power supply limitations.  Actuation
typically takes the form of I/O (e.g., network, peripherals,
displays).

At the OS/runtime level, the computation specification relation,
$f \subseteq \mathbb{I} \times \mathbb{O}$ takes the form of
guarantees provided by the system. These may be guarantees and the
resulting definition of correctness in terms of the numeric behavior
of the computation, or may be guarantees on timeliness of operations
in real-time and interactive computing systems. At this layer,
faults and errors typically refer to the failure of a component
from the architecture level and its manifestation in a difference
in machine state respectively.

\vspace{-0.05in}
\subsection{Runtime systems: computation}
\vspace{-0.05in}
Trading timeliness guarantees for reduced resource usage was heavily
explored in the 1990s, in research efforts on \textit{imprecise
realtime systems}~\cite{hull1993ics, liu1991algorithms, shih1995algorithms,
aydin99incorporating, liu:impreciseoverview}. Much like the recent
resurgence of interest in trading correctness for resource usage,
these earlier efforts were targeted at an application domain (embedded
systems) where the relaxation of correctness requirements was
motivated by the inherent nondeterminism of their operating
environments.

%
%

The Eon system~\cite{Sorber2007} provides a declarative language
which allows users to annotate components with different energy
specifications, which are then used at runtime to select suitable
components. Hoffman\etalii~\cite{Hoffmann2011} turn static configuration
parameters into dynamic knobs which can adjust the accuracy and
energy usage of a system at runtime. A calibration pass offline
minimizes monitoring overhead at runtime. Similarly, Green~\cite{Baek2010}
builds a quality of service model during a calibration phase based
on approximations supplied by the programmer. This model is used
at runtime to select suitable approximations.
Chippa\etalii~\cite{Chippa2011} propose a general framework which
phrases the dynamic management as a feedback system and further
suggest different quality measurements at the circuit, architecture,
and algorithm level which serve as the feedback signal.

Most previous runtime approaches consider average errors or only
check the errors occasionally during execution, and can thus miss
large outliers. Rumba~\cite{Khudia2015} checks all results with
light-weight checks and proposes an approximate correction mechanism,
which is specific to data-parallel applications. Topaz~\cite{Achour2015}
also verifies every result, but at a higher
granularity, by decomposing a computation into tasks. Topaz checks each task's
output with lightweight checks provided by the user.  If
the error is too large, Topaz automatically re-executes the
corresponding task.

\vspace{-0.05in}
\subsection{Runtime systems: sensors, actuation, and displays}
\vspace{-0.05in}
All measurements have some amount of measurement uncertainty and
as a result, sensing systems provide many opportunities for trading
errors for improved efficiency. These range from trading accuracy
and reliability in sensors in the Lax system~\cite{189934}, to
trading precision for fidelity in imaging sensors
(cameras)~\cite{buckler-iccv2017}, to trading the fidelity of display
colors and shapes for reduced display panel power dissipation for
OLED displays in Crayon~\cite{Stanley-Marbell:2016:CSP:2901318.2901347}.

\vspace{-0.05in}
\subsection{Summary}
\vspace{-0.05in}
OS and runtime techniques provide a unique opportunity to exploit
dynamic information about running programs. Unlike circuit-level,
microarchitectural, architectural, or language-level techniques,
they can exploit information about a user's environment such as
remaining energy store in a mobile device or activation of a low-power
mode on the device. OS and runtime techniques also have the opportunity
to learn across program executions. Hardware platforms for exploring
the end-to-end benefits of the techniques presented in this survey
(Challenges 2 and 8 in Section~\ref{section:challenges}) may however
be necessary for a meaningful evaluation of real-world benefits.

%
%
\vspace{-0.05in}
\section{Taxonomy}
\label{section:taxonomy}
\vspace{-0.05in}
Table~\ref{table:taxonomy} highlights techniques for trading
correctness for resource usage discussed throughout this survey.
The table focuses on publications that present a specific technique
as opposed to publications discussed in the survey to provide context.
Table~\ref{table:taxonomy} classifies techniques by three primary
categories: (1) \emph{error type}, (2) \emph{property traded for
errors}, and (3) \emph{affected resources}:
\begin{itemize} 
\item \textbf{The error type} refers to the nature of the error
that gets introduced into a system.  Given the same input and set
of initial conditions, a technique is deterministic if it will
always cause the same outcome and a technique is nondeterministic
if the outcome can differ.

\item \textbf{The property traded for errors} is one of \emph{energy},
\emph{time}, and \emph{data density}.  These are cost functions
that a system designer optimizes for. In the context of
this survey, we consider trading an improvement in one or more of
these properties for increased occurrence of errors.

\item \textbf{The affected resources} are the hardware
subsystems that are impacted by the tradeoffs. In practice, these
will be the subsystems in which errors occur.
\end{itemize}

%
%

\definecolor{tablegray}{gray}{0.92}
\definecolor{tablewhite}{gray}{1.00}

\newcolumntype{w}{>{\columncolor{tablegray}}l}
\newcolumntype{x}{>{\columncolor{tablewhite}}l}
\newcolumntype{y}{>{\columncolor{tablegray}}c}
\newcolumntype{z}{>{\columncolor{tablewhite}}l}

\begin{table*}
\caption{Highlights from the techniques covered in this survey,
from circuit-level techniques, to architecture-level techniques,
to algorithmic and programming-language-level techniques, to operating
system techniques. We classify the techniques under the three broad
categories of (1) \textit{error type}, (2) \textit{property traded for
errors}, and (3) \textit{affected resources}. (PL: Programming language; OS: Operating system.)}
\vspace{-0.1in}
\scriptsize 
\centering
\begin{tabular}{@{\extracolsep{-5pt}}xwxyyzzzyyy}
\toprule
\textbf{Layer}	& \textbf{Technique}	& \textbf{Examples}	& \textbf{Error} & \textbf{Type~} & \textbf{Property} & \textbf{Traded} & \textbf{for} & \textbf{Affected} & \textbf{Reso} & \textbf{\!\!\!\!urce}\\
& & & & & \textbf{Errors} & & & & & \\
\midrule
& & & & & & & & & & \\
& & & & & & & & & & \\
& & & & & & & & & & \\
& & & & & & & & & & \\
& & & & & & & & & & \\
& & & & & & & & & & \\
& & & & & & & & & & \\
& & & & & & & & & & \\
                 &                       &                      & \begin{rotate}{90}{\emph{Deterministic}}\end{rotate} & \begin{rotate}{90}{\emph{Non-Deterministic}}\end{rotate} & \begin{rotate}{90}{\emph{Energy}}\end{rotate} & \begin{rotate}{90}{\emph{Runtime}}\end{rotate} & \begin{rotate}{90}{\emph{Data Density}}\end{rotate}           & \begin{rotate}{90}{\emph{Computation}}\end{rotate} & \begin{rotate}{90}{\emph{Data~Storage/Movement}}\end{rotate} & \begin{rotate}{90}{\emph{Physical~World~~I/O}}\end{rotate} \\
\midrule
\emph{Circuit} 
            & Sensor value approximation & \cite{buckler-iccv2017,189934}    &           & $\bullet$ & $\bullet$ &           &           &           &           & $\bullet$ \\
            & Probabilistic sensor comms. & \cite{1804.02317}           & $\bullet$ &           & $\bullet$ &           &           &           &           & $\bullet$ \\
            & Probabilistic computing      & \cite{cct:p1,cct:p2,cct:p3,cct:p4,cct:p5}  && $\bullet$ & $\bullet$ &     &   &  $\bullet$        &        &  \\
            & Stochastic computing       & \cite{Brown2001, Qian11, Alaghi14}& $\bullet$ &           &           & $\bullet$ &           &  $\bullet$ &           &           \\
            & Voltage overscaling  & \cite{cct:shanhbag:sdsp,cct:roy:vos,cct:Kahng:vos,cct:kurdahi:vos,cct:TERRA}&           & $\bullet$ & $\bullet$ & & &  $\bullet$  &           &  \\
            & Logic pruning              & \cite{cct:p7,cct:ls1,cct:ALS}     & $\bullet$ &           & $\bullet$ & $\bullet$  &  $\bullet$ &  $\bullet$  &           &  \\
            & Approximate addition  & \cite{cct:AMA,cct:LOA,cct:ETA,cct:Dith,cct:config1}& $\bullet$ &  & $\bullet$ & $\bullet$  & $\bullet$ &  $\bullet$  &           &  \\
            & Approximate multiplication & \cite{cct:Mul1,cct:Mul2,cct:Mul3} & $\bullet$ &           & $\bullet$ & $\bullet$  & $\bullet$ &  $\bullet$  &           &  \\
            & RTL approximations & \cite{cct:SALSA,cct:ABACUS}               & $\bullet$ &           & $\bullet$ & $\bullet$  & $\bullet$ &  $\bullet$  &           &  \\
            & Approx.\ high-level synthesis & \cite{cct:AHLS,cct:AHLS2}&       $\bullet$ &           & $\bullet$ & $\bullet$  & $\bullet$ &  $\bullet$  &           &  \\
            & Voltage overscaled SRAM  & \cite{cct:SRAM,cct:hybrid1,cct:hybrid2,cct:cache}                 &     &  $\bullet$   & $\bullet$ &    & &          & $\bullet$       &  \\
\midrule
\emph{Architecture~~~~~~}& Deterministic lossy I/O& \cite{Stanley-Marbell:2016:RSI:2897937.2898079, hotchips16encoder, kim2017axserbus}
                                                                             & $\bullet$ &           & $\bullet$ &           &           &           & $\bullet$ &           \\
            & Voltage overscaling          & \cite{esmaeilzadeh:architecture}&           & $\bullet$ & $\bullet$ &           &           & $\bullet$ & $\bullet$ &           \\
            & Analog neural acceleration   & \cite{amant:general}            &           & $\bullet$ & $\bullet$ & $\bullet$ &           & $\bullet$ &           &           \\
            & Digital neural acceleration  & \cite{esmaeilzadeh:neural, yazdanbakhsh:neural, moreau:snnap}
                                                                             & $\bullet$ &           & $\bullet$ & $\bullet$ &           & $\bullet$ &           &           \\
            & Anytime computation          & \cite{sanmiguel:anytime}        & $\bullet$ &           &           & $\bullet$ &           & $\bullet$ &           &           \\
            & Approximate reads            & \cite{ranjan:approximate}       &           & $\bullet$ & $\bullet$ & $\bullet$ &           &           & $\bullet$ &           \\
            & Approximate writes           & \cite{sampson:approximate}      &           & $\bullet$ & $\bullet$ & $\bullet$ &           &           & $\bullet$ &           \\
            & Reuse of failed data blocks  & \cite{sampson:approximate}      &           & $\bullet$ &           &           & $\bullet$ &           & $\bullet$ &           \\
            & Variable redundancy          & \cite{jevdjic:approximate, guo:high, fujiki:highbandwidth}
                                                                             &           & $\bullet$ &           &           & $\bullet$ &           & $\bullet$ &           \\
            & Approx. cache de-duplication  & \cite{sanmiguel:bunker,sanmiguel:doppelganger}
                                                                             & $\bullet$ &           &           &           & $\bullet$ &           & $\bullet$ &           \\
            & Load-value approximation     & \cite{sanmiguel:load, thwaites:rollback}
                                                                             & $\bullet$ &           & $\bullet$ & $\bullet$ &           &           & $\bullet$ &           \\
            & Low-refresh DRAM             & \cite{liu:flikker}              &           & $\bullet$ & $\bullet$ &           &           &           & $\bullet$ &           \\
            & In-network lossy compression & \cite{boyapati:approxnoc}       & $\bullet$ &           &           &           & $\bullet$ &           & $\bullet$ &           \\
\midrule
\emph{PL}   & Floating-point optimization  & \cite{Chiang2017,Darulova2018,Rubio-Gonzalez2013}
                                                                             & $\bullet$ &           &           & $\bullet$ &           & $\bullet$ &           &           \\
            & Neural approximation         & \cite{esmaeilzadeh:architecture}& $\bullet$ &          & $\bullet$ & $\bullet$ &           & $\bullet$ &           &           \\
            & Relaxed parallelization      & \cite{rinard2006,Misailovic2012,Misailovic2013,Renganarayana2012,campanoni2015}
                                                                             &           & $\bullet$ &           & $\bullet$ &           & $\bullet$ &           &           \\
            & Accuracy verification        & \cite{carbin2012,Misailovic2014}           &           & $\bullet$ & $\bullet$ &           &           & $\bullet$ &           &           \\
            & Data-parallel kernel approx. & \cite{Samadi2014}               & $\bullet$ &           &           & $\bullet$ &           & $\bullet$ &           &           \\
            & Isolation of approx. data    & \cite{Sampson2011,Park2015}     & $\bullet$ & $\bullet$ & $\bullet$ &           &           & $\bullet$ & $\bullet$ &           \\
            & Algorithm approximation      & \cite{Schkufza2014}             & $\bullet$ &           &           & $\bullet$ &           & $\bullet$ &           &           \\
            & Loop perforation             & \cite{Sidiroglou-Douskos2011}   & $\bullet$ &           &           & $\bullet$ &           & $\bullet$ &           &           \\ 
\midrule
\emph{OS}   & Display color approximation  & \cite{Stanley-Marbell:2016:CSP:2901318.2901347}
                                                                             & $\bullet$ &           & $\bullet$ &           &           &           &           & $\bullet$ \\
            & Drivers for approx. sensors  & \cite{189934}
                                                                             &           & $\bullet$ & $\bullet$ &           &           &           &           & $\bullet$ \\
            & Dynamic accuracy adaptation  & \cite{Sorber2007, Hoffmann2011, Baek2010} 
                                                                             & $\bullet$ &           & $\bullet$ &           &           & $\bullet$ &           &           \\
            & Task-level approximation     & \cite{Achour2015} 
                                                                             &           & $\bullet$ & $\bullet$ &           &           & $\bullet$ &           &           \\
\bottomrule
\end{tabular}
\normalsize
\label{table:taxonomy}
\end{table*}

Although the table places publications in discrete categories, many
techniques lie somewhere in a spectrum. For example, when voltage
overscaling (Section~\ref{section:05:overscaling}) is performed at
a coarse level (e.g., in steps of 500\,mV for a device with a supply
voltage range of 1.8\,V to 3.3\,V), it could be seen as a deterministic
technique where some voltage levels always lead to repeatable
failures. On the other hand, if voltage overscaling is performed at
a fine granularity of voltages (e.g., 50\,mV), there will likely
be one or more voltage levels where nondeterministic failures
occur, resulting from the interplay between devices operating right
at the threshold of the minimum voltage for reliable operation, and
falling below that threshold due to power supply noise or thermal
noise.

At the circuit level, most techniques in the research literature
to date have focused on trading errors for energy efficiency and
to a lesser extent, performance and data storage density. At this
level of the system stack, the focus has been overwhelmingly on computation
resources (e.g., arithmetic/logic units (ALUs)) as the \textit{Affected
Resources} columns in Table~\ref{table:taxonomy} show.

Most of the architectural techniques in Table~\ref{table:taxonomy}
target computation at a coarse grain (e.g., analog and digital
neural accelerators). A majority of the architectural techniques
listed in Table~\ref{table:taxonomy} apply to data movement and
storage such as on- and off-chip memories, memory hierarchy data
traffic, and on- and off-chip I/O links.

Programming language techniques have largely focused only on
techniques that affect computation, as the \textit{Affected Resources}
columns in Table~\ref{table:taxonomy} show. This is unsurprising,
since most programming languages focus on providing primitives and
abstractions for computations (as opposed to, say, communication).
There is potentially an unexplored opportunity to investigate
techniques for trading errors for efficiency applied to language-level
constructs for communication such as the channels in Hoare's
communicating sequential processes (CSP).  One early investigation
of this direction is the M language, which provided language-level
error, erasure, and latency tolerance constraints~\cite{M:pmup06}
on CSP-style language-level channels.

Techniques implemented at the operating system (OS) level, such as
application programming interfaces (APIs), standard system
libraries, device drivers, and so on, have the unique vantage point
of seeing all system processes. Techniques at the OS-level often
have a global view of the system hardware, and visibility into
application behavior beyond a single execution instance. OS-level
techniques also have access to information about user preferences,
such as activation of a low-power mode on a mobile device. Despite
these potential advantages of OS-level techniques for trading errors
for efficiency, there have been relatively few techniques implemented
at this level of abstraction. The techniques which
Table~\ref{table:taxonomy} lists target improving energy efficiency
and do so primarily by trading the use of sensors, displays, and
coarse-grained application level error behavior for improved
efficiency.

\vspace{-0.05in}
\section{Fundamental Limits of Resource versus Correctness Tradeoffs}
\label{section:fundamentalLimitsOfNoisyComputation}
\vspace{-0.05in}
Section~\ref{section:transistorGateAndCircuitLevel} through
Section~\ref{section:OsLevel} presented concrete techniques for
trading resource usage for correctness at levels of abstraction
ranging form the device-, gate-, and circuit-level, to the operating
system. For techniques at each of these levels of abstraction, this
article formulated the resource usage versus correctness tradeoff
in terms of a computational problem, its implementation in an
algorithm, and a distance function $d$ between the state representations
of a computation's correct and resource-reduced variants. That
relation between a computation's input and output or between a
computation's state prior to and subsequent to computation has
parallels to communication systems. We can draw an analogy between
the state transformation performed by an algorithm which must consume
resources (time, energy, die area) to achieve the exact behavior
specified by the computational problem which it implements, and
source- and channel-coding for communication over a channel: Source-
and channel-coding which likewise consume resources in order to
maximize the mutual information between the transmitter and receiver
over a channel. Von Neumann~\cite{VonNeumann:problogic}, Berger and
Gibson~\cite{BergerGibson:98}, Evans~\cite{evans1998maximum},
Maggs~\cite{Cole:rfgarrays}, Elias~\cite{elias1958computation},
Spielman~\cite{spielman1996highly}, and Shanbhag~\cite{HShanbhag:ANT},
among others, have previously drawn similar analogies between
resource usage versus correctness tradeoffs and communication
channels. And doing so provides a useful lens through which to study
the fundamental limits of resource usage versus correctness tradeoffs
in computing systems.

%
%
%


\vspace{-0.05in}
\subsection{From information and coding theory to coded computation}
\vspace{-0.05in}
The study of fault-tolerant systems dates back to von Neumann's
investigation~\cite{VonNeumann:problogic} of building reliable
systems from unreliable components. Fault-tolerant systems research
has focused more heavily on a coarser-grained view. In contrast,
\textit{information theory} focuses on the mathematical study of
communication over noisy channels~\cite{Shannon:59} while \textit{coding
theory} studies methods for judiciously trading
redundancy in data representations for either reduced transmission
time (source coding) or improved end-to-end reliability in transmission
over a noisy channel (channel coding).

In contrast to channel coding techniques whose objective is to
counteract the effect of noise, Chen\etalii~\cite{chen2014noise}
exploit the presence of noise to improve image processing tasks,
demonstrating how adding Gaussian noise to quantized images can
improve the output quality of signal processing tasks. This observation
that noise can improve a computing system's performance has parallels
to randomized algorithms (see, e.g., Section~\ref{section:taxonomy}).

%
%
%

Classical information and coding theory rely on the assumption that noise
only occurs in communication, rather than in computation.  In
contrast, recent research has begun to study the fundamental limits
of encoders~\cite{yang2014can} and decoders~\cite{varshney2011performance,
yazdi2013gallager} built on top of hardware implementations that
are, like the communication channel, susceptible to noise. Similarly,
recent research has investigated techniques for executing computation
on encoded representations in order to obtain exact or approximate
results in the presence of noise. These methods have been referred
to in the research literature as \textit{coded
computation}~\cite{grover2014shannon, rachlin2008framework}.
One plausible direction for future research is
to identify computing abstractions that unify the above techniques
via new computational operators that execute on encoded representations.
Stochastic computing~\cite{cct:SC}, hyper-dimensional computing~\cite{Rabaey2017},
and deep embedded representations (deep learning) offer promising
examples.

%
%
%

%
%
%

\vspace{-0.05in}
\subsection{Theoretical bounds}
\vspace{-0.05in}
Recent research has used information theory as a foundation to
investigate theoretical bounds on performance~\cite{yazdi2012optimal},
efficiency~\cite{stanley2009encoding}, energy
consumption~\cite{chatterjee2016energy}, Shannon-style channel
capacity and storage bounds~\cite{stanley2009encoding,
varshney2011performance} for computing and communication systems
which trade resource usage for correctness.
Varshney~\cite{varshney2011performance} demonstrates
Shannon-style bounds on storage capacity in the context of noisy
LDPC iterative decoders.  Stanley-Marbell~\cite{stanley2009encoding}
derives best-case efficiency bounds for encoding techniques which
limit the deviations of values in the presence of logic upsets.
Chatterjee\etalii~\cite{chatterjee2016energy} present lower bounds
on energy consumption for achieving a desired level of reliability
in computation of an $n$-input Boolean function and
Yazdi\etalii~\cite{yazdi2012optimal} formulate an optimization
problem to produce a noisy Gallager B LDPC decoder that achieves
minimal bit error rate, by treating unreliable hardware components
as communication channels as in stochastic computing (see
Section~\ref{section:transistorGateAndCircuitLevel} for coverage
of stochastic computing).  These recent research efforts demonstrate
that information and coding theory can provide a baseline to derive
bounds on efficiency, capacity, energy consumption, and performance
in the systems of interest in this survey: computing systems which
trade resource usage for correctness.


\vspace{-0.05in}
\subsection{Application-aware source and channel coding across the hardware stack}
\vspace{-0.05in}
Mitigating the effects of errors across the stack will
ultimately require encoding techniques, applied across the layers
of the stack that are designed to take advantage of application
characteristics.  Early examples of such \emph{application-aware
codes} can be found in the work of Huang\etalii~\cite{huang2015acoco}
which proposes a redundancy-free adaptive code that can correct
errors in data retrieved from potentially faulty cells.  The technique
relies on an application-specific cost function and an \emph{input-adaptive
coding scheme} that pairs a source encoder that introduces modest
distortion, with a channel encoder that adds redundant bits to
protect the distorted data against errors.  Adaptive coding can
greatly reduce output quality degradation in the presence of noise,
compared to na\"ive implementations
where noise is allowed to traverse the stack unchecked.

\vspace{-0.05in}
\subsection{Summary}
\vspace{-0.05in}
Information and coding theory today form the basis for techniques
to analyze and model noisy communication and storage systems as
well as techniques to counteract the effects of noise. 
With the emergence of approximate computing, there is an opportunity to investigate
new approaches to source and channel coding.
These new approaches
could explicitly take into account the specific noise distributions
observed in practice and could explicitly take into account the
requirements of the applications consuming the data in question.
These new challenges require a unifying mathematical theory to
reason about errors, efficiency, and capacity bounds.

%
%


\vspace{-0.05in}
\section{Challenges}
\label{section:challenges}
\vspace{-0.05in}

%
%
We identify eight challenges facing both research and applications
of techniques to trade correctness for resource usage.

%
%

\noindent\textbf{Challenge 1: Holistic cross-layer approaches.}\quad A
whole-system view to trading errors for efficiency requires expertise
in the target application domain and in multiple levels of the
computing stack. Most of the existing approximation and error-handling
mechanisms are designed in the context of a single layer in the
stack. This is likely to be suboptimal. Techniques in different
layers can easily negate each other, where gains reported in isolation
may not translate into real system-level benefits in the end. At
the same time, techniques in different layers may complement each
other, where significant opportunities for cross-layer optimizations
can be expected. A full-stack view of error-efficient system design
requires less insular approaches. A cross-layer approach will however significantly
increase the size of the design space and could introduce significant additional design
complexity.

\noindent\textbf{Challenge 2: Hardware models, hardware platforms, and measurement
data.}\quad Most soft-ware-level techniques employ models or abstractions
of the errors and performance of the underlying hardware in order
to achieve modularity and scalability.  Examples of hardware error
models assumed today include assumptions about the distribution of
locations and values of errors caused by voltage overscaling in
microarchitectural structures and memories, or assumptions about
the distribution of errors in DRAMs that are not refreshed as
regularly as they should be. Similarly, lower levels of the software
stack may expose higher-level models to, e.g., application
developers. Today, different research efforts often use different
models, which makes comparisons between research results difficult
and raises questions about the validity of reported resource savings.
Error and performance models which have been validated by the
hardware community, e.g., by hardware measurements and which are
suitable for the software levels of the stack would be an invaluable
contribution to the research directions described in this survey.
In order to make credible claims about across-the-stack approximations,
the proposed techniques need to be evaluated end-to-end either on
actual state-of-the-art hardware platforms or with realistic
simulations.  Such end-to-end evaluations with an agreed-upon
platform is missing today.  Early
examples in this direction include measurement results from open
hardware platforms explicitly designed to expose accuracy, precision,
performance, and energy efficiency tradeoffs~\cite{1804.09241}.

\noindent\textbf{Challenge 3: Hardware emulation/simulation, software tools,
languages, and compiler infrastructure.}\quad Applying error versus
resource tradeoffs in software requires tools that help programmers
and systems builders take advantage of techniques in a systematic
and controlled way. It also requires hardware simulators or emulators
that help bridge the gap between the fidelity of hardware
prototypes and the flexibility of software simulation. On the
hardware simulation side, these tools would ideally provide support
for end-to-end evaluation of entire systems, to be used in
comparing different proposed techniques. Language and compiler tools
would include those to support testing, debugging, and dynamic
quality monitoring. First steps in this direction for compiler tools
include ACCEPT~\cite{sampson2015accept}.

\noindent\textbf{Challenge 4: Application domains and algorithmic patterns.}\quad
Today, there is insufficient consensus on well-defined classes of
application domains and algorithmic patterns that constitute a
preferential target for relaxations.  First steps include the
definition of ``Recognition, Mining, and Synthesis'' application
classes~\cite{dubey2005recognition}.  A standard benchmark set which
has been agreed upon by the wider community would increase progress
and comparability of different techniques, like it has done for
SAT/SMT solving. Such a benchmark set should ideally cover different
domains and include also real-world `challenge applications' which
cannot be solved today, but which would convincingly demonstrate
the viability of error-efficient computing.

\noindent\textbf{Challenge 5: Large-scale user studies to provide empirical
characterization of acceptability.}\quad User studies with thousands of
participants will be necessary to provide quantitative
data~\cite{lockhart2011design}, which researchers can use when
proposing techniques that exploit tolerance of human observers to
deviations from correctness of program results. Initial steps in
this direction include the ``Specimen'' dataset of color perception
data used in color approximation
techniques~\cite{DBLP:journals/corr/abs-1803-08420}.

\noindent\textbf{Challenge 6: Metrics.}\quad When applying techniques to an application,
it would be useful to have a reliable error metric to guide the
optimization process.  In the ideal case, that error metric would
be binary: ``correct'' and ``not correct.'' But in reality, correctness
and its boundaries are not well known for many applications. The
metrics in question might be broadly applicable to many systems,
or might be application-specific metrics used to measure
Pareto-optimality. Early work in this direction includes the work of Akturk\etalii~\cite{akturk2015quantification}.

\noindent\textbf{Challenge 7: Studies of the theoretical bounds on resource
usage.}\quad Theoretical upper and lower bounds are invaluable in guiding
research as they set the bar for what is achievable. Such bounds
are needed for a given formally-defined specification of deviation
from correctness. One promising direction are bounds on encoding
overhead for communication encoding techniques which trade communication
energy use or performance (data rate) for integer deviations in
communicated values. Examples of first steps in this direction
include work on the bounds of encoding efficiency for deterministic
and probabilistic approximate communication
techniques~\cite{stanley2015efficiency, stanley2009encoding}.

\noindent\textbf{Challenge 8: Reproducibility and deployment of techniques.}\quad
This survey describes a broad range of techniques, from circuits
to algorithms. For many of the research results across these layers,
it can be challenging to replicate the setup, tools, or benchmarks
employed in evaluations. Beyond good scientific practice of describing
experiments in sufficient detail to be reproducible, because there
is today no common consensus even on many aspects of terminology,
it is challenging to compare, replicate, and build upon research
results.  This survey attempted to address the challenge of terminology
with a consistent set of formal definitions across the layers of a
system (Section~\ref{section:terminology}). Further progress is
however needed. Opportunities include greater availability of
open-source libraries. First steps for low-power and high-performance
approximate arithmetic components include synthesizable Verilog/VHDL files
and behavioral models in C/MATLAB~\cite{Shafique15,
shafique2016invited}.

		%
		%
		%
		%

\vspace{-0.05in}
\section{Retrospective and Future Directions}
\label{section:conclusions}
\vspace{-0.05in}
Computing systems use resources such as time, energy, and hardware
to transform their inputs to outputs. For many years, the primary
driver of efficiency improvements in computing were a combination
of improved semiconductor process technology and better algorithms.
In the last decade, two important shifts have forced a fundamental
re-evaluation of the hardware driver of efficiency improvements.
First, with the cessation of constant-field Dennard scaling and the
stagnation of device supply voltages, process technology scaling
no longer provides the improvements in energy efficiency that it
once did. Second, in contrast to traditional computing applications
such as financial transaction processing and office productivity,
the dominant computing system applications are increasingly driven
by inputs from the physical world (audio, images) with outputs for
consumption by humans.  Applications driven by data from the physical
world have essentially unbounded input datasets, and this has partly
motivated a resurgence of interest in machine learning approaches
to learning functions from large datasets. The stagnation in hardware
device-level improvements coupled with increasingly ever more
abundant sensor-data-centric workloads has led to a need for new
ways of improving computing system performance.
This survey explored techniques to address this challenge of computing
on data when less-than-perfect outputs are acceptable for a computing
system's users.



\vspace{-0.05in}
\section*{Acknowledgements}
\vspace{-0.05in}
This article is the result of a four-day workshop held in April
2017 organized by Phillip Stanley-Marbell, Adrian Sampson, and Babak
Falsafi. Many of the ideas in the survey resulted from discussions
among the entire group of participants (in alphabetical order):
Sara Achour, Armin Alaghi, Mattia Cacciotti, Michael Carbin,
Alexandros Daglis, Eva Darulova, Lara Dolecek, Mario Drumond, Natalie
Enright Jerger, Babak Falsafi, Andreas Gerstlauer, Ghayoor Gillani,
Djordje Jevdjic, Sasa Misailovic, Thierry Moreau, Adrian Sampson,
Phillip Stanley-Marbell, Radha Venkatagiri, Naveen Verma, Marina
Zapater, and Damien Zufferey. The workshop was funded by the Swiss
National Science Foundation (SNF) under grant  \emph{IZ32Z0\_173393/1:
``International Exploratory Workshop on Theory and Practice for
Error-Efficient Computing''}, and by the EPFL EcoCloud Research
Center. Ghayoor Gillani thanks A.B.J. Kokkeler (CAES group, University
of Twente) for his encouragement and suggestions to contribute
towards this article.

\bibliographystyle{ACM-Reference-Format}
\bibliography{survey}

\end{document}